\documentclass[
a4paper,
aps,
prd,
preprint,
english,
tightenlines,
preprintnumbers,
nofootinbib,
floatfix
]{revtex4-1}

\usepackage[utf8]{inputenc}
\usepackage{csquotes}
\MakeOuterQuote{"}
\usepackage[kerning=true]{microtype}
\usepackage{braket}
\usepackage{fullpage}
\usepackage{amsfonts}
\usepackage{amsmath}
\usepackage{amssymb} 
\usepackage{graphicx}
\usepackage{epic}
\usepackage{eepic}
\usepackage{epsfig}
\usepackage{latexsym}
\usepackage[usenames,dvipsnames,svgnames]{xcolor} %Provides colors
\usepackage{color}
\usepackage{float}
\usepackage{multirow} 
\usepackage[caption=false]{subfig}
\usepackage{slashed}

\usepackage{graphicx}
\usepackage{booktabs}
\usepackage{multirow}
\definecolor{darkteal}{HTML}{1B9E77}	
\definecolor{darkorange}{HTML}{D95F02}
\definecolor{darklilac}{HTML}{7570B3}
\definecolor{darkmagenta}{HTML}{E7298A}
\definecolor{darklimegreen}{HTML}{66A61E}
\definecolor{darkbanana}{HTML}{E6AB02}
\definecolor{darktan}{HTML}{A6761D}
\definecolor{darkgray}{HTML}{666666}

\newcommand{\iu}{{i\mkern1mu}}
\newcommand{\room}{~}
\usepackage{shorthand}

\usepackage[colorlinks=true,allcolors=darkmagenta]{hyperref} 

\begin{document}

%\preprint{UU-XXXXYYYY}

\title{Minimal anomalous $\mathrm{U}(1)$ theories and collider phenomenology}

\author{Andreas Ekstedt}
\email{andreas.ekstedt@physics.uu.se}
\affiliation{Department of Physics and Astronomy, Uppsala University, Box 516, SE-751 20 Uppsala, Sweden}

\author{Rikard Enberg}
\email{rikard.enberg@physics.uu.se}
\affiliation{Department of Physics and Astronomy, Uppsala University, Box 516, SE-751 20 Uppsala, Sweden}

\author{Gunnar Ingelman}
\email{gunnar.ingelman@physics.uu.se}
\affiliation{Department of Physics and Astronomy, Uppsala University, Box 516, SE-751 20 Uppsala, Sweden}

\author{Johan L\"ofgren}
\email{johan.lofgren@physics.uu.se}
\affiliation{Department of Physics and Astronomy, Uppsala University, Box 516, SE-751 20 Uppsala, Sweden}

\author{Tanumoy Mandal}
\email{tanumoy.mandal@physics.uu.se}
\affiliation{Department of Physics and Astronomy, Uppsala University, Box 516, SE-751 20 Uppsala, Sweden}
\affiliation{Department of Physics and Astrophysics, University of Delhi, Delhi 110007, India}

%\date{\today}

\begin{abstract} 
We study the collider phenomenology of a neutral gauge boson $Z'$ arising in
minimal but anomalous $\mathrm{U}(1)$ extensions of the Standard Model (SM).
To retain gauge invariance of physical observables, we consider cancellation of
gauge anomalies through the Green-Schwarz mechanism. We categorize a wide
class of $\mathrm{U}(1)$ extensions in terms of the new $\mathrm{U}(1)$
charges of the left-handed quarks and leptons and the Higgs doublet. We 
derive constraints on some benchmark models using electroweak precision 
constraints and the latest 13 TeV LHC dilepton and dijet resonance search
data. We calculate the decay rates of the exotic and rare one-loop $Z'$ decays to 
$ZZ$ and $Z$-photon modes, which are the unique signatures of our framework. If observed,
these decays could hint at anomaly cancellation through the
Green-Schwarz mechanism. We also discuss the possible observation of such 
signatures at the LHC and at future ILC colliders. 
\end{abstract}

%\pacs{}

%\keywords{}

\maketitle

\section{Introduction}\label{sec_intro}
The prospect of discovering a heavy and neutral gauge boson, often dubbed $ Z' $, at the LHC has motivated many different phenomenological studies of models in which such particles arise. A simple example of such a model is a $\mathrm{U}(1)$-extension of the standard model (SM). If one wishes to consider $\mathrm{U}(1)$-models with chiral fermions in a consistent manner, one should take care that gauge-invariance is not violated by anomalies. In order to enforce this, traditionally one constructs the classical action of the theory to be gauge-invariant, together with choosing particular relations between the gauge charges of the chiral fermions such that the anomalies cancel\room\cite{Appelquist:2002mw}. For a recent update of collider bounds on such models, see \cite{Ekstedt:2016wyi} and references therein.

However, this is not the only possible way to enforce gauge-invariance. An alternative is to consider the possibility of adding gauge-variant terms to the classical Lagrangian such that the full theory with anomalies satisfies all Ward identities. By accepting this point of view it is possible to abandon the notion that the classical action has to be gauge-invariant, and consider a theory which has gauge-dependent building blocks but obeys all relevant Ward identities in the end. This idea can be realized through the Green-Schwarz (GS) mechanism\room\cite{Green:1984sg} which can arise in several different settings, e.g., in string theories, or from integrating out heavy fermions.\footnote{Here we are being a bit cavalier with the term gauge-invariance. It should be noted that one really deals with a gauge-fixed Lagrangian, for which BRST invariance is the remaining symmetry that the observables must obey.}

The principal idea is this\room\cite{Antoniadis:2010zza}: Gauge-invariance should be apparent at all energies -- even if anomaly cancellation is taken care of by high-scale physics. Thus, the contribution of such physics, e.g., heavy fermions running in loops, \emph{no matter how heavy}, should not be suppressed at low energies. In\room\cite{Anastasopoulos:2006cz} the authors conclude that such an effective action and its phenomenological consequences cannot determine the nature of the high-scale physics. Even though this conclusion does not offer an additional window into high-scale physics, it does allow fairly model-independent studies of the GS mechanism. With this in mind, we will in this paper perform a more detailed phenomenological analysis (aimed primarily at the LHC) of GS $\mathrm{U}(1)$ extensions. For earlier phenomenological work in this setting, see\room\cite{Anastasopoulos:2008jt} for a pre-LHC analysis of an extension of the MSSM,\room\cite{Ismail:2017ulg} for a more recent collider study in the context of explaining dark matter, and\room\cite{Dror:2017ehi,Dror:2017nsg,Ismail:2017fgq} for studies where the anomalous $ Z' $ is very light. The assumptions of our approach include, (i) an additional $\mathrm{U}(1)$ gauge group broken by the St\"{u}ckelberg mechanism, (ii) SM fermions are the only fermions (not integrated out) which are charged under the SM gauge group, (iii) the gauge charges are generation independent, and (iv) the electroweak symmetry breaking (EWSB) occurs as in the SM.

In section\room\ref{sec:u1extn}, we discuss minimal $ \mathrm{U}(1) $-extensions of the SM, with focus on the GS mechanism in subsection\room\ref{ssec:GS}. In section\room\ref{sec:models}, we describe various interesting models which are possible in this setting. We describe the computations of branching ratios, including details regarding the evaluation of 1-loop processes, in section\room\ref{sec:BR}. In section\room\ref{sec:results}, we review our phenomenological results, capped off with a discussion in section\room\ref{sec:discussion}.

\section{Minimal $\mathbf{U(1)}$ extensions}\label{sec:u1extn}

We consider a generic $\mathrm{U}(1)$ extension of the SM whose gauge group 
is $\mathrm{SU}(3)_C \times \mathrm{SU}(2)_L \times \mathrm{U}(1)_Y \times \mathrm{U}(1)_z$. The gauge couplings, gauge fields and field strengths
associated with $\{\mathrm{SU}(3)_C,\mathrm{SU}(2)_L,\mathrm{U}(1)_Y,\mathrm{U}(1)_z\}$ are $\{g_S,g,g',g_z\}$; $\{G^{\mu},W^{\mu},B^{\mu}_Y,B^{\mu}_z\}$; and $\{F_G^{\mu\nu},F_W^{\mu\nu},F_Y^{\mu\nu},F_z^{\mu\nu}\}$, respectively. In this paper, we consider anomaly cancellation via the GS mechanism, and the extra Abelian $\mathrm{U}(1)_z$ is broken to the SM gauge group at some high scale through the St\"{u}ckelberg mechanism, which makes the $Z'$ massive. The EWSB then proceeds as usual; the details of the symmetry breaking can be found in appendix\room\ref{app:ssb}.

The Higgs doublet $\Phi$ and all the SM fermions are in general all charged under
$\mathrm{U}(1)_z$. The three generations of left-handed quark and lepton doublets are denoted by 
$q_L^i$ and $l_L^i$ respectively and the right-handed components of up-type, down-type quarks and
charged leptons are denoted by $u_R^i$, $d_R^i$ and $e_R^i$ (here $i=1,2,3$)
respectively.
We denote the hypercharge by $Y$ and the $\mathrm{U}(1)_z$ charge by $z$, which we assume to be generation independent to prevent flavor changing neutral currents. The charges of the different particles are labeled according to the convention of\room\cite{Appelquist:2002mw}, which is summarized in table\room\ref{tab:charges}. The $\mathrm{U}(1)_z$ charges of fermions are constrained 
to provide gauge-invariant Yukawa couplings, i.e. 
$z_{u}=z_{q}+z_{H};~z_{d}=z_{q}-z_{H};~z_{e}=z_{\ell}-z_{H}$. 
\begin{table}
\centering
\begin{tabular}{c|c|c|c|c}
%\hline
Fields & $\mathrm{SU}(3)_{c}$ & $\mathrm{SU}(2)_{L}$ & $\mathrm{U}(1)_{Y}$ & $\mathrm{U}(1)_{z}$ \\
\hline
$H$ & 1 & 2 & $1$ & $z_H$\\
$q_{L}$ & 3 & 2 & $1/3$ & $z_{q}$ \\
$u_{R}$ & 3 & 1 & $4/3$ & $z_{u}=z_{q}+z_H$\\
$d_{R}$ & 3 & 1 & $-2/3$ & $z_{d}=z_{q}-z_{H}$\\
$\ell_{L}$ & 1 & 2 & $-1$ & $z_\ell$\\
$e_{R}$ & 1 & 1 & $-2$ & $z_{e}=z_\ell-z_H$\\
%\hline
\end{tabular}
\caption{The $\mathrm{U}(1)_{z}$ charge assignments of the Higgs doublet and the fermions of the SM.}\label{tab:charges}	
\end{table}

\subsection{Anomaly cancellation and $\mathbf{U(1)_z}$ charges}\label{ssec:charges}

In order to construct an anomaly-free gauge theory with chiral fermions, it is common to assign the gauge charges of the fermions such that the gauge anomalies cancel when the contributions from all fermions are taken into account. 
Gauge anomalies are always proportional to a trace over all relevant fermions.
Introduction of a $\mathrm{U}(1)_z$ symmetry leads to six types of possible anomalies, which are shown in table\room\ref{tab:anomalies} together with the corresponding traces and their expressions in terms of the free charges $z_q,z_\ell$ and $z_H$ (these expressions are similar to the ones derived in\room\cite{Anastasopoulos:2008jt}). This table also includes the corresponding GS parameters for future reference. It should be noted that the mixed gauge anomaly $\lt[\mathrm{SU}(3)_c\rt]^2\lt[\mathrm{U}(1)_z\rt]$ cancels automatically when the Yukawa coupling constraints are enforced.

\begin{table}
\centering
\begin{tabular}{cccc}
Anomaly&Trace&Parameters&Expression\\
\hline
$\lt[\mathrm{U}(1)_z\rt]^3$ & $\textrm{Tr}\lt[z^3\rt]$ & $C_{zzz}$ & $-z_H^3 - 3 z_H z_\ell^2 - z_\ell^3 + 3 z_H^2 (z_\ell + 6 z_q)$ \\
$\lt[\mathrm{U}(1)_z\rt]^2\lt[\mathrm{U}(1)_Y\rt]$ & $\textrm{Tr}\lt[Yz^2\rt]$&$E_{zzy},C_{zzy}$&$4 z_H (z_\ell + 3 z_q)$\\
$\lt[\mathrm{U}(1)_z\rt]\lt[\mathrm{U}(1)_Y\rt]^2$ &$\textrm{Tr}\lt[Y^2z\rt]$ & $E_{zyy},C_{zyy}$ & $4 (z_\ell + 3 z_q)$\\
$\lt[\mathrm{SU}(2)_L\rt]^2\lt[\mathrm{U}(1)_z\rt]$\hspace{0.5em} &$\textrm{Tr}\lt[\lt\{T^i,T^j\rt\}z\rt]$ & $K_{2},D_{2}$ & $(6 z_q+2 z_\ell)$\\
$\lt[\mathrm{SU}(3)_C\rt]^2\lt[\mathrm{U}(1)_z\rt]$\hspace{0.5em} & $\textrm{Tr}\lt[\lt\{\mc{T}^a,\mc{T}^b\rt\}z\rt]$& $K_{3},D_{3}$ & $0$\\
$\lt[\mathrm{R}\rt]^2\lt[\mathrm{U}(1)_z\rt]$& $\textrm{Tr}\lt[z\rt]$ &--- & $3 z_q+2 z_\ell-z_{H}$ 
\end{tabular}
\caption{The different possible gauge anomalies, together with the corresponding traces, the corresponding GS parameters and the traces' algebraic expressions in terms of the charges $z_q,z_\ell$ and $z_H$. The table is expressed in terms of the generators $T^i$ of $\mathrm{SU}(2)_L$, $\mc{T}^a$ of $\mathrm{SU}(3)_c$, $Y$ of $\mathrm{U}(1)_Y$ and $z$ of $\mathrm{U}(1)_z$. In the final anomaly we have written R to represent the general relativity gauge group.}\label{tab:anomalies}
\end{table}

If the anomalies are canceled via the appropriate fermion charge assignments, the general solution to the anomaly cancellation conditions (in the framework with no kinetic mixing) is for the charge $Q_{f}^{z}$ of a given fermion $f$ under the gauge group $ \mathrm{U}\left(1\right)_{z} $ to be written as a linear combination of its hypercharge $Y_{f}$ and $(B-L)_{f}$ quantum number~\cite{WeinbergII}, i.e., $Q_{f}^{z}=aY_{f}+b(B-L)_{f}$. However, if the charges are "free", the most general fermion charge can be written in terms of $z_q,~z_\ell$ and $z_H$ as
\begin{equation}\label{eq:charge}
Q_{f}^{z}=3z_{q} B_{f}+z_{\ell} L_{f} +z_{H} \left\{ Y_{f}-\left( B-L\right)_{f}  \right\}.
\end{equation}

\subsection{The Green-Schwarz mechanism}\label{ssec:GS}

In this subsection, we review how the GS mechanism\room\cite{Green:1984sg} can be used to generate a low-energy effective action which is anomaly free -- for a more formal review of gauge anomalies, 
see~\cite{Adler:1969er,Adler:1969gk,Fujikawa:1979ay,Fujikawa:1980eg}.

Anomalies associated with the $\mathrm{U}(1)_z$ extensions are, in general, both mixed and pure. Pure anomalies only violate BRST symmetry for particular gauge transformations, while mixed anomalies introduce violation of multiple transformations. An anomaly is called \emph{relevant} if it is not possible to completely remove it by adding a local counterterm to the classical Lagrangian. However, it is always possible, by reshuffling the mixed anomalies, to put all anomalous transformations on the $\mathrm{U}(1)_z$ group. Explicitly, if we integrate out all fermions we can define an effective action as
\begin{align*}
e^{\iu \Gamma}=\int \prod_{\text{fermions}}\mathcal{D}\Phi e^{i\mathcal{S}}.
\end{align*}
A typical mixed $\mathrm{U}(1)$ anomaly has the form\room\cite{Bilal:2008qx},
$\delta\Gamma\sim A \, \theta_Y \epsilon_{\alpha \beta \mu \nu}  F_z^{\alpha \beta}F_z^{\mu \nu}+B\,\theta_z \epsilon_{\alpha \beta \mu \nu}F_z^{\alpha \beta} F_Y^{\mu \nu}$, where $ \theta_Y $ and $ \theta_z $ are the gauge transformation parameters of the respective $ \mathrm{U}(1) $ groups.
Adding a counterterm $\mathcal{L}_\text{ct}\sim A \epsilon_{\alpha \beta \mu \nu} B_Y^\alpha B_z^\beta F_z^{\mu \nu}$ alters the anomalous transformation to
\begin{align*}
\delta\Gamma\rightarrow \left(A+B\right)\theta_z \epsilon_{\alpha \beta \mu \nu}F_z^{\alpha \beta} F_Y^{\mu \nu}.
\end{align*}
For \emph{relevant} anomalies, it is not possible to completely remove the remaining 
$\mathrm{U}(1)_z$ anomaly with the available field content. However, since all the $\mathrm{U}(1)_z$ anomalous transformations are of the form $\sim \theta_z \mathrm{Tr}(F^2)$, it is possible to add a pseudoscalar, $\mathcal{A}$, to the spectrum, transforming under $\mathrm{U}(1)_z$ as $\mathcal{A}\rightarrow \mathcal{A}+M g_z \theta_z $. The anomalous $\mathrm{U}(1)_z$ transformation can then be removed by adding terms of the form 
$\mathcal{L}\sim (\mathcal{A}/M)\mathrm{Tr}(F^2)$ to the Lagrangian. This is a low-energy form of the GS mechanism.

In a $\mathrm{U}(1)$ extension of the SM, the GS mechanism can be incorporated by using the formalism developed in\room\cite{Anastasopoulos:2006cz}. Three types of new terms appear in the Lagrangian,
\begin{equation}
\label{eq:LagGS}
\mathcal{L} \supset \mathcal{L}_\text{kin}+\mathcal{L}_\text{PQ}+\mathcal{L}_\text{GCS}.
\end{equation}
The first term $\mathcal{L}_\text{kin}$ consists of kinetic energy terms of the $\mathrm{U}(1)_z$ gauge boson $B_z$ together with the pseudoscalar $\mathcal{A}$ (also known as a St\"{u}ckelberg axion \cite{Stueckelberg:1900zz}) as follows,
\begin{equation}
\mathcal{L}_\text{kin} = -\frac{1}{4} \left(F_z^{\mu\nu}\right)^2 
+ \frac{1}{2}\left(\partial^{\mu}\mathcal{A}+M g_z B_z^{\mu} \right)^2 ,
\end{equation}
where $F_z^{\mu\nu}$ is the field strength tensor of the $B_z^{\mu}$ field and $M$ is a parameter with the dimension of mass, further discussed at the end of this subsection. The kinetic terms are chosen such that $\mathcal{L}_\text{kin}$ is invariant under the $\mathrm{U}(1)_z$ transformation 
$B_z^{\mu} \to B_z^{\mu} - \partial^{\mu}\theta_z$ and $\mathcal{A} \to \mathcal{A} + M g_z \theta_z$.
The second and third parts of Eq.~\eqref{eq:LagGS}, $\mathcal{L}_\text{PQ}$ and $\mathcal{L}_\text{GCS}$,  are called the \emph{Peccei-Quinn} (PQ) 
 and the \emph{generalized Chern-Simons} (GCS)  terms 
respectively. These two classes of terms, as described above, are chosen such that they remove all gauge anomalies. 

The Lagrangian $\mathcal{L}_\text{PQ}$ contains couplings between $\mc{A}$ and gauge-invariant terms of the form $\mathrm{Tr}(F^2)$, in a fashion similar to the PQ mechanism \cite{Peccei:1977hh},
\begin{align}\label{eq:LagPQ}
\mathcal{L}_\text{PQ}=\frac{\hbar}{16 \pi^2}\frac{1}{6 M} \mc{A}~\varepsilon_{\mu\nu\rho\sigma}\left(\right.&C_{zzz} g_z^2 F_z^{\mu\nu} F_z^{\rho\sigma}+C_{zzy} g_z g' F_z^{\mu\nu} F_Y^{\rho\sigma}+C_{zyy} g'^2 F_Y^{\mu\nu} F_Y^{\rho\sigma}\nn\\
+& D_{2} g^2 \mathrm{Tr}\left(F_W^{\mu\nu} F_W^{\rho\sigma}\right)+D_{3} g_S^2 \mathrm{Tr}\left(F_S^{\mu\nu} F_S^{\rho\sigma}\right)\left.\right).
\end{align}

The $\mathcal{L}_\text{GCS}$ part is chosen such that its gauge transformations mimic the mixed anomalies, and contains antisymmetric trilinear interactions of various gauge bosons. These
can be written as
\begin{align}\label{eq:LagGCS}
\mathcal{L}_\text{GCS}=\frac{\hbar}{16 \pi^2}\frac{1}{3 }\varepsilon_{\mu\nu\rho\sigma}\left(\vphantom{\Omega_S^{\nu\rho\sigma}}\right.&g'^2 g_z E_{zyy} B_Y^\mu B_z^\nu F_Y^{\rho\sigma}+g' g_z^2 E_{zzy} B_Y^\mu B_z^\nu F_z^{\rho\sigma}\nn\\
+ &g^2 g_z K_{2} B_z^\mu  \Omega_W^{\nu\rho\sigma}+g_S^2 g_z K_{3} B_z^\mu\Omega_S^{\nu\rho\sigma}  \left.\right),
\end{align}
where $\Omega$ is the non-Abelian Chern-Simons 3-form (here we write $A^{S},A^{W}$ instead of $G,W$ to simplify the notation), given by
\begin{equation}\label{eq:CS}
\Omega^{S,W}_{\nu\rho\sigma}=\frac{1}{3}\mathrm{Tr}\left[A^{S,W}_{\nu}\left(F^{S,W}_{\rho\sigma}-[A^{S,W}_{\rho},A^{S,W}_{\sigma}]\right)+\left(\mathrm{cyclic~perm.}\right)\right].
\end{equation}
In equations\room\eqref{eq:LagPQ} and\room\eqref{eq:LagGCS} we have restored a factor of $ \hbar $, to emphasize that these terms are of 1-loop strength. The various coefficients ($C,D,E,K$) in equations~\eqref{eq:LagPQ} and~\eqref{eq:LagGCS} can be expressed in terms of the different $\mathrm{U}(1)_z$ charges of the fermions by matching the new terms' transformation to the anomalies\room\cite{Anastasopoulos:2006cz}.
\begin{align}
	C_{zzz} &=-\frac{3}{8}\left(z_h^3 + 3 z_h z_\ell^2 + z_\ell^3 - 3 z_h^2 (z_\ell + 6 z_q)\right)\label{eq:czzz},\\
	C_{zzy} &= -\frac{9}{2} z_h (z_\ell + 3 z_q)=3 E_{zzy}\label{eq:czzy},\\
	C_{zyy} &=-\frac{9}{4} (z_\ell + 3 z_q) =\frac{3}{2} E_{zyy}\label{eq:czyy},\\
	D_{2} &=\frac{9}{2}(6 z_q+2 z_\ell) =-\frac{3}{2} K_{2}\label{eq:D2},\\
	D_{3} &=0=K_{3}\label{eq:D3}.
\end{align} 
The coefficients $D_3,K_3$ are zero due to the fact that the $\lt[\mathrm{SU}(3)_c\rt]^2\lt[\mathrm{U}(1)_z\rt]$ anomaly cancels automatically from the gauge invariance of the Yukawa sector.

From the $\mathcal{L}_\text{PQ}$ terms we see that this theory contains vertices including axion and gauge bosons of the form $\mc{A}ZZ,\mc{A}Z'Z',\mc{A}\gamma\gamma,\mc{A}W^{+}W^{-}$ (the coupling to gluons is zero). The $\mathcal{L}_\text{GCS}$ part generates the new tree-level vertices 
$ZZ\gamma,ZZ'\gamma,Z'Z'\gamma$, which are not present in traditional anomaly-free 
$\mathrm{U}(1) $-extensions\room\cite{Appelquist:2002mw}. As described above, these new terms serve, in practice, as counter-terms for anomalous amplitudes, as for example the standard triangle-fermion amplitude. We are especially interested in the amplitudes $Z' ZZ$ and $Z' \gamma Z$, which, if observed, may give indications of the GS nature of the theory.

The parameter $ M $ introduced in $ \mathcal{L}_\text{kin} $ and $ \mathcal{L}_\text{PQ} $ has the dimension of mass and corresponds to a high scale. It can be interpreted as a vacuum expectation value of a Higgs field which spontaneously breaks $ \mathrm{U}(1)_{z} $. If the corresponding physical scalar is heavy, it can be integrated out. The remnant is a pseudoscalar boson $ \mc{A} $ and a mass term for the $ B_{z} $ field. In our minimal setup $ \mc{A} $ is not physical, but simply a Goldstone boson which is absorbed by the gauge fields. By considering a more complicated Higgs sector it is possible to furnish a physical axion and a Goldstone boson, through mixing with other scalar fields.

Note that the PQ terms are suppressed by the scale $ M $; from equation\room\eqref{eq:scale} in appendix\room\ref{app:ssb} it can be seen that $ M\sim M_{Z'}/g_{z} $ as $ M_{Z'}\rightarrow\infty $.
However, the GCS terms remain unsuppressed even at low energies, see Eq.~(\ref{eq:LagGCS}).

\subsection{Ward identities in the broken theory}

In perturbative calculations, gauge anomalies manifest as the violations of various Ward identities for both the unbroken and the broken theory. A case relevant for
the $Z'$ phenomenology is the process $Z'\rightarrow \gamma Z$, which should obey the Ward identities
\begin{align}
&p^\mu_{Z'}\Gamma^{Z'\gamma Z}_{\mu \nu \rho}-\iu M_{Z'}\Gamma^{\phi_{Z'}\gamma Z}_{\nu \rho}=0,
\\&p^\nu_{\gamma}\Gamma^{Z'\gamma Z}_{\mu \nu \rho}=0,
\\&p^\rho_{Z}\Gamma^{Z'\gamma Z}_{\mu \nu \rho}-\iu M_{Z}\Gamma^{Z' \gamma \phi_{Z}}_{\mu \nu}=0,
\end{align}
where, e.g., $ \Gamma^{Z'\gamma Z}_{\mu\nu\rho} $ is the amputated 
$ Z'^{\mu}\gamma^{\nu}Z^{\rho}$ three-point function with all momenta outgoing and $\phi_{Z},\phi_{Z'}$ denote the Goldstone bosons corresponding to $Z,Z'$ respectively. Anomalies present in the unbroken theory will be inherited in the broken theory, and show up as violations of the Ward identities for the spontaneously broken theory.

In the example above, the Ward identities will be broken by terms proportional to the $\left[\mathrm{U}(1)_Y\right]^2 \left[\mathrm{U}(1)_z\right]$ and $\left[\mathrm{SU}(2)_L\right]^2 \left[\mathrm{U}(1)_z\right]$ anomalies (together with the relevant mixing angles). 
In addition, a process such as $Z'\rightarrow Z Z$ would also inherit the anomaly $\left[\mathrm{U}(1)_z\right]^3$. This anomaly is not present in the $Z'\to\gamma Z$ case since the photon does not mix with the $Z'$. An easy way to see this is to recall that right-handed neutrinos are often introduced to cancel the $\left[\mathrm{U}(1)_z\right]^3$ anomaly (and the gravity anomaly), and at the lowest order calculation of $Z'\rightarrow \gamma Z$, right-handed neutrinos cannot circulate in the fermion loop since they do not couple to the photon.

For a concrete example consider one of the anomalous $Z' Z Z$ Ward identities that takes the form
\begin{align}
p^\mu_{Z'}\Gamma^{Z' Z Z}_{\mu \nu \rho}- iM_{Z'}\Gamma^{\phi_{Z'} Z Z}_{\nu \rho}=i\epsilon_{\nu \rho\alpha \beta}p_\gamma^\alpha p_Z^\beta\lt(\frac{g_z A}{96 \pi^2 c_w^2s_w^2}\rt),
\end{align}
where $c_w=\cos\theta_W$ and $s_w=\sin\theta_W$ ($\theta_W$ is the Weinberg angle). The anomalous factor $A$, in the model considered above, is given by
\begin{align}
A &= \left(c_z^2 - 2s_z^2\right)\left\lbrace\right. 2 e^2 c_z s_w^4 \left(2 z_d+6 z_e-3 z_\ell - z_q+8 z_u\right)\nn\\
&- \left. 6 e^2 c_z c_w^4 \left(z_\ell+3z_q\right) - 6 e g_z s_z c_w s_w^3 \left(z_d^2+z_3^2-z_\ell^2+z_q^2-2z_u^2\right)\right\rbrace\nn\\
&-3g_z^2 c_z s_z^2 s_w^2 c_w^2\left(3 z_d^3+z_e^3-2 z_\ell^3-6 z_q^3+3 z_u^3\right),
\end{align}
where $c_z=\cos\theta'$ and $s_z=\sin\theta'$ ($\theta'$ is the $Z\leftrightarrow Z'$ mixing angle).
The origin of each term is clear; for instance, the presence of 
$\cos^4\theta_W$ and $\sin^4\theta_W$ indicate that these terms correspond to the 
$\left[\mathrm{SU}(2)_L\right]^2 \left[\mathrm{U}(1)_z\right]$ and $\left[\mathrm{U}(1)_Y\right]^2 \left[\mathrm{U}(1)_z\right]$  anomalies, respectively. While the terms related with $\left[\mathrm{U}(1)_z\right]^2 \left[\mathrm{U}(1)_Y\right]$ and $\left[\mathrm{U}(1)_z\right]^3$ anomalies come with extra factors of $g_z$ and $\sin\theta'$, since they are absent if 
$\theta'=0$. The different GS terms are hence constructed to cancel these anomalous terms.

\section{Interesting models}\label{sec:models}

In Eq.~\eqref{eq:charge}, we can see that the $\mathrm{U}(1)_{z}$ charge of a given fermion can be written in terms of the free charges $z_q,z_\ell$ and $z_H$ and the quantum numbers $ Y,B $ and $ L $. Note that the form of the charge is completely determined by the spontaneous symmetry breaking together with the assumption of generation-independent charges. With this charge known, it is now possible to consider different interesting models. First, there are the traditional models described in, e.g., Ref.~\cite{Appelquist:2002mw}, which we will not consider in this paper (a popular example is gauged $B-L$ models). In the GS setting, however, there are more exotic possibilities. We divide them into two categories: chiral (C) and non-chiral (NC) models. Since hypercharge is the only chiral charge present in~\eqref{eq:charge}, the NC models are categorized by $z_{H}=0$. All of the NC models correspond to different linear combinations of $B$ and $L$. Here is a list of examples:
\begin{itemize}
	\item $Q_{f}^{z}=B_{f}$ (baryon number): Obtained by choosing the charges $ z_{q}=1/3;z_{\ell}=0;z_{H}=0$. This model is leptophobic\room\cite{Carone:1995pu}.
	\item $Q_{f}^{z}=L_{f}$ (lepton number): $ z_{q}=0;z_{\ell}=1;z_{H}=0$. This model is quarkphobic\room\cite{Aranda:2014zta}.
	\item $Q_{f}^{z}=B_{f}-L_{f}$: $ z_{q}=1/3;z_{\ell}=-1;z_{H}=0$. This is a widely studied traditional model\room\cite{Mohapatra:1982xz} which can be made anomaly free by including right-handed neutrinos.
	\item $ Q_{f}^{z}=0 $ (fermiophobic): $ z_{q}=0;z_{\ell}=0;z_{H}=0$. This model is anomaly free trivially.
	\item $Q_{f}^{z}=B_{f}+L_{f}$: $ z_{q}=1/3;z_{\ell}=1;z_{H}=0$.
	\item $Q_{f}^{z}=z_{\ell}\left(L_{f}-2B_{f}\right)$: $ z_{q}=-(2/3)z_{\ell};z_{H}=0;$ with $ z_{\ell} $ free. This model is an NC example of the gravity model (see the list of C models below). 
\end{itemize}
For C models, we need $ z_{H}\neq 0$. A list of examples is
\begin{itemize}
	\item $ Q_{f}^{z}=z_{H}Y_{f} $ (Y-sequential): Obtained by choosing the charges $ z_{q}=(1/3)z_{H};z_{\ell}=-z_{H};$ with $ z_{H} $ free but nonzero. This model is automatically anomaly free since it is just a copy of the SM $ \mathrm{U}(1)_{Y} $ gauge group\room\cite{Erler:1999ub}. 
	\item $ Q_{f}^{z}=-(1/2)\left(B-L\right)_{f}+(1/5)Y_{f} $ ($\mathrm{SO}(10)$ GUT): 
	$z_{q}=-1/10; z_{\ell}=3/10;z_{H}=1/5$. This model can be made anomaly free by adding right-handed neutrinos\room\cite{DeRujula:1980qc}.
	\item $Q_{f}^{z}=z_{H}\left( Y_{f}-\left( B-L\right)_{f}  \right)$ (right-handed): $ z_{q}=0;z_{\ell}=0;$ with $ z_{H} $ free but nonzero. With $ z_{H}=-1/2 $ one obtains the traditional right-handed model which can be made anomaly free by adding right-handed neutrinos\room\cite{Erler:1999ub}.
	\item $ Q_{f}^{z}= z_{H}\left(Y_{f}-B_{f}\right)+\left(z_{\ell}+z_{H}\right)L_{f}$ (right-handed quarks): $z_{q}=0;$ with $ z_{\ell} $ free and $ z_{H} $ free but nonzero.
	\item $ Q_{f}^{z}= 3z_{q}B_{f}+z_{\ell}\left(2L_{f}+Y_{f}\right)$ (left-handed leptons): $z_{h}=z_{\ell};$ with $ z_{q} $ free and $ z_{\ell} $ free but nonzero.
	\item $ Q_{f}^{z}= \left(3z_{q}-z_{H}\right)B_{f}+z_{H}\left(Y_{f}+L_{f}\right)$ (right-handed leptons): $z_{\ell}=0$ with $ z_{q} $ free and $ z_{H} $ free but nonzero.
	\item $ Q_{f}^{z}= \left(3z_{q}+(1/2)z_{\ell}\right)B_{f}+(1/2)z_{\ell}\left(L_{f}-Y_{f}\right)$ (axial leptons): $z_{H}=-(1/2)z_{\ell}$ with $ z_{q} $ free and $ z_{\ell} $ free but nonzero.			
	\item $Q_{f}^{z}=-2z_{\ell}B_{f} +3(z_{q}+z_{\ell})L_{f}+(3z_{q}+2z_{\ell})Y_{f}$: $ z_{H}=(3z_{q}+2z_{\ell})$ with $ z_{q},z_{\ell} $ free such that $ z_{\ell}\neq -(3/2) z_{q}  $. This model is constructed to cancel the gauge-gravity anomaly explicitly.
\end{itemize}
In this paper, we will focus on four benchmark models, but we also perform random scans of the parameter space of charges. Note that the models which are automatically anomaly free have all the GS parameters equal to zero. The models which can be made anomaly free by adding right-handed neutrinos have many of the GS parameters equal to zero, but not all of them, and hence have weak exotic signatures. We note that all of these models necessarily have $ z_{\ell}=-3 z_{q} $.

\section{$Z'$ decays and partial widths}\label{sec:BR}

In our models, $Z'$ has the following tree-level decays: $Z'\to\bar{f}f$ (where $f$ denotes any SM fermion), 
$Z'\to W^+W^-$, and $Z'\to ZH$. There are also two possible one-loop decays of $Z'$,
$Z'\to Z\gm$ and $Z'\to ZZ$, whereas the $Z'\to\gm\gm$ decay is forbidden by the Landau-Yang theorem. Although the branching ratios (BRs) of these loop-suppressed decay modes are very small, they can act
as unique signatures of the GS mechanism and are hence of particular interest. The analytical formulas of the tree level two-body decay modes are easy to compute and are given in\room\cite{Ekstedt:2016wyi}. The production cross-sections are calculated in the \textsc{Madgraph} package\room\cite{Alwall:2014hca}. The loop level decay modes have been calculated using the \textsc{FeynCalc} package\room\cite{Shtabovenko:2016sxi,Mertig:1990an}, with the Feynman rules calculated using \textsc{FeynRules}\room\cite{Alloul:2013bka}, and the diagrams generated using \textsc{FeynArts}\room\cite{Kublbeck:1990xc,Hahn:2000kx}. To evaluate the one loop integrals we use \textsc{Package-X}\room\cite{Patel:2015tea}, which is interfaced to \textsc{FeynCalc} by \textsc{FeynHelpers}\room\cite{Shtabovenko:2016whf}. Details of these calculations can be found in subsection\room\ref{ssec:Loop} below, and in appendix\room\ref{app:Loops}.

\subsection{Loop induced decays}\label{ssec:Loop}
Both of the $Z'\rightarrow Z \gamma$ and $Z'\rightarrow Z Z$ processes are of loop strength and do not appear at the tree level. These processes are interesting since they receive contributions from the GS terms, and could indicate the presence of such terms.  Both of the above-mentioned processes are finite and contain a gauge contribution and a fermionic contribution, but it turns out that the gauge loops cancel (see \cite{Anastasopoulos:2008jt}) and only the fermion loops are non-zero. We calculate these processes in the symmetric anomaly scheme\room\cite{Anastasopoulos:2006cz} and evaluate them in the limit where all fermion masses excluding the top mass vanish. The other fermion masses give negligible contributions to the amplitude.\footnote{A similar analysis of anomalous amplitudes was performed in\room\cite{Ismail:2017ulg}.} The notation $\epsilon[\mu,\nu,\rho,q]\equiv \epsilon[\mu,\nu,\rho,\alpha]q^\alpha$ will be used extensively. 
\begin{figure}[h]
\centering
\includegraphics[width=0.6\textwidth]{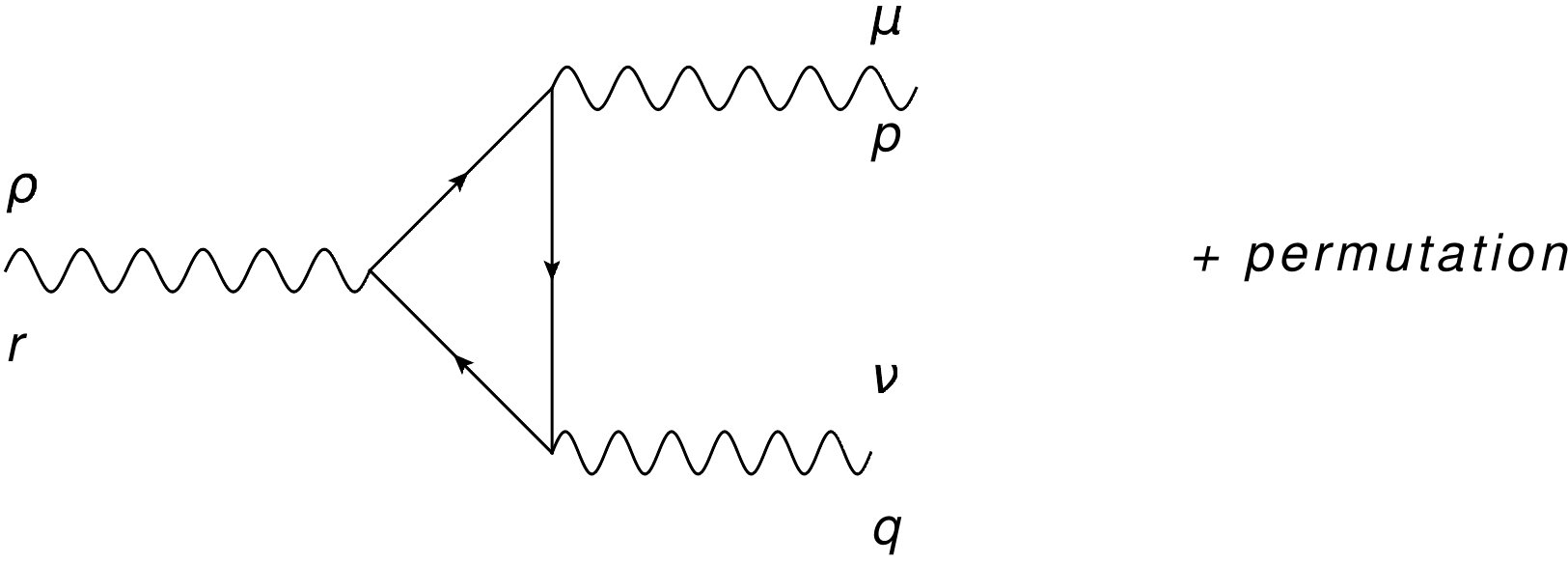}
\caption{Generic fermion loop for $Z'$ decay into two vector bosons.}\label{fig:triangle}
\end{figure}

\subsubsection{$Z' \to ZZ$ decay}\label{sssec:ZpZZ}
The triangle loop for this process is shown in fig.\room\ref{fig:triangle}; the amplitude is denoted as
\begin{align}
\Gamma^{Z' Z Z}_{\rho \mu \nu}(r,p,q),
\end{align}
where the $Z'$ momentum $r=p+q$ is incoming, the $Z$ momenta $p ,q$ are outgoing, and $ p^{2}=q^{2}=M_{Z}^{2} $. The generic process can be parametrized as
\begin{align}
\Gamma^{Z' Z Z}_{\rho \mu \nu}(r,p,q) &= A_1~ \epsilon[\mu ,\nu,p,q]q^\rho+ A_2~  \epsilon[\mu ,\nu,p,q]p^\rho\nonumber
\\&+ A_3 ~\epsilon[\mu,\nu, \rho,q]+ A_4 ~\epsilon[\mu,\nu, \rho,p]\nonumber
\\& + A_5 ~\epsilon[\nu,\rho,p,q]q^\mu+ A_6 ~\epsilon[\nu,\rho,p,q]p^\mu\nonumber
\\& + A_7 ~\epsilon[\mu,\rho,p,q]q^\nu+ A_8 ~\epsilon[\mu,\rho,p,q]p^\nu,
\end{align}
where $ A_1 $--$ A_8 $ are Lorentz-invariant functions of $ p,q $ and $m_f$ (see appendix\room\ref{app:Loops} for the explicit forms of these functions). Bose symmetry, i.e., symmetry under the replacements ($\mu \leftrightarrow \nu, p \leftrightarrow q$), dictates $A_1=A_2,~A_3=-A_4,~A_5=-A_8, A_6=-A_7$. In addition, the relations $A_5=-A_6,~A_7=-A_8$ hold, which can be seen after applying the relevant Ward identities. The amplitude contribution from a single fermion can hence be written in the compact form
\begin{align}\label{eq:zpzzPA}
\Gamma^{Z' Z Z}_{\rho \mu \nu}(r,p,q) &= A \left( \epsilon[\mu ,\nu,p,q]q^\rho+\epsilon[\mu ,\nu,p,q]p^\rho\right)\nonumber
\\&+ B\left(\epsilon[\mu,\nu, \rho,q] - \epsilon[\mu,\nu, \rho,p]\right)\nonumber
\\& + C\left( \epsilon[\nu,\rho,p,q]q^\mu-\epsilon[\nu,\rho,p,q]p^\mu+\left(\mu \leftrightarrow \nu\right)\right).
%\\& +C \left(\epsilon[\mu,\rho,p,q]q^\nu- \epsilon[\mu,\rho,p,q]p^\nu\right).
\end{align}
It is possible to rewrite the amplitude in the Rosenberg parametrization\room\cite{Rosenberg:1962pp} by using the Schouten identity (see appendix\room\ref{app:Loops} for details).
The complete transition amplitude can then be written as
\begin{align}
\mathcal{T}^{Z' Z Z}_{\lambda' \lambda_1 \lambda_2}(r,p,q)=\epsilon^{\lambda'}(r)_\rho\epsilon^{\lambda_1}_\mu(p)\epsilon^{\lambda_2}(q)_\nu \Gamma^{Z' Z Z}_{\rho \mu \nu}(r,p,q),
\end{align}
where the A terms, shown in equation\room\eqref{eq:zpzzPA}, drop out from the calculation due to the transversality of the polarization tensors. Averaging over initial state polarization and summing over final state polarizations the square of the complete amplitude takes the form
\begin{align}
\braket{\left|\mathcal{T}\right|^{2}} & \equiv  \frac{1}{3}\sum_{\lambda',\lambda_1,\lambda_2=\pm,0}\mathcal{T}^{Z' Z Z}_{\lambda' \lambda_1 \lambda_2}(r,p,q) \left(\mathcal{T}^{Z' Z Z}_{\lambda' \lambda_1 \lambda_2}(r,p,q)\right)^{*}\nonumber
\\  & = \frac{\left(M_{Z'}^2-4 M_{Z}^2\right)^2}{12 M_{Z}^2}\left| \sum_f \left(2 B_f+ M_{Z'}^2 C_f\right)\right|^2,
\end{align}
where $B_f,~C_f$ denote the form factor contributions in equation (\ref{eq:zpzzPA}) for a specific fermion $f$, and a sum over all fermions has been included. The GCS-terms will have the same Lorentz structure as the $B$-term; including these in the amplitude gives
\begin{align}
\braket{\left|\mathcal{T}\right|^{2}}= \frac{\left(M_{Z'}^2-4 M_{Z}^2\right)^2}{12 M_{Z}^2}\left| \sum_f \left(2 B_f+ M_{Z'}^2 C_f\right)+2 (\text{GCS})^{Z' Z Z}\right|^2.
\end{align}

The decay width is then given by
\begin{align}
\Gamma^{Z' Z Z}=\frac{1}{2}\frac{1}{16 \pi M_{Z'}} \sqrt{1 - 4 \left(\frac{M_{Z}}{M_{Z'}}\right)^2} \braket{\left|\mathcal{T}\right|^{2}},
\end{align}
where the the symmetry factor has been included due to identical final states.

The role of the GCS terms can best be seen in the $M_{Z'}\rightarrow \infty$ limit, in which the form factor simplifies to $2 B_f+M_{Z'}^2 C_f\rightarrow 2 B_f|_{M_{Z'}\rightarrow  \infty }$, where the leading order term $\sum_f 2 B_f|_{M_{Z'}\rightarrow  \infty }$ is mass independent and proportional to the anomaly $ \mathcal{A} $: $B_f=\mathcal{A}+\mathcal{O}\left(\frac{1}{M_{Z'}^3}\right)$. The GCS terms cancel the leading order term, $\text{GCS}=-\sum_f 2 B_f|_{M_{Z'}\rightarrow  \infty }$. This cancellation ensures that the process is unitary.

\subsubsection{$Z' \to Z \gamma$ decay}\label{sssec:ZpZg}

The $Z'\rightarrow Z \gamma$ amplitude is shown in fig.\room\ref{fig:triangle}, and the evaluation is very similar as for the $Z' \rightarrow Z Z$ process. The amplitude is denoted as
\begin{align}
\Gamma^{Z' Z \gamma}_{\rho \mu \nu}(r,p,q),
\end{align}
where $r=p+q$, and $q^2=0,~p^2=M_{Z}^2$. The amplitude can be written as
\begin{align}
\Gamma^{Z' Z \gamma}_{\rho \mu \nu}(r,p,q) &= A_1~ \epsilon[\mu ,\nu,p,q]q^\rho+ A_2~  \epsilon[\mu ,\nu,p,q]p^\rho\nonumber
\\&+ A_3 ~\epsilon[\mu,\nu, \rho,q]+ A_4 ~\epsilon[\mu,\nu, \rho,p]\nonumber
\\& + A_5 ~\epsilon[\nu,\rho,p,q]q^\mu+ A_6 ~\epsilon[\nu,\rho,p,q]p^\mu\nonumber
\\& + A_7 ~\epsilon[\mu,\rho,p,q]q^\nu+ A_8 ~\epsilon[\mu,\rho,p,q]p^\nu.
\end{align}
In contrast with the $Z'\rightarrow Z \gamma$ amplitude, there are no direct Bose-symmetry relations, but it still turns out that the decay width is completely characterized by two form factors. Using the Schouten identity for light-like momenta (see appendix \ref{app:Loops}),
\begin{align}
q^\rho \epsilon[\mu,\nu,p,q]=-q^\mu \epsilon[\nu, \rho,p,q]+p\cdot q \epsilon[\mu,\nu,\rho,q],
\end{align}
together with transversality of the polarization tensors, we can exchange $A_2$ for $-A_1$ and remove $A_6$ and $A_7$. This leaves the Lorentz structure
\begin{align}
\Gamma^{Z' Z \gamma}_{\rho \mu \nu}(r,p,q) &= B_1 ~\epsilon[\mu,\nu, \rho,q]+ B_2 ~\epsilon[\mu,\nu, \rho,p]\nonumber
\\& + B_3 ~\epsilon[\nu,\rho,p,q]q^\mu+ B_4 ~\epsilon[\mu,\rho,p,q]p^\nu.
\end{align}
The above functions are not all independent, which can be seen by using the three Ward identities of the amplitude, or from the explicit calculations in appendix \ref{app:Loops}. The remaining form factors are related as: 
	\begin{align*}
	B_2 &= p \cdot q B_3-\mathcal{A},
	\\ B_3 &=- B_4,
	\\ B_1 &=-B_3 (p\cdot q- M_{Z}^2) \\
	   &-\frac{3\iu Q_f m_f^2 (g_{Z,f}^R-g_{Z,f}^L)(g_{Z',f}^R-g_{Z',f}^L) C_0(0,M_{Z}^2,M_{Z'}^2,m_f^2,m_f^2,m_f^2)}{12 \pi^2}+\mathcal{A},
	\end{align*}%
where $\mathcal{A}$ is a combination of the anomaly terms and the contribution from the GCS terms; $ g_{Z',f}^{R} $ and $ g_{Z',f}^{L} $ are the right-handed and left-handed couplings respectively; $C_0(0,M_{Z}^2,M_{Z'}^2,m_f^2,m_f^2,m_f^2)$ is the usual Passarino-Veltman scalar integral.

The above relations leave two independent form factors, such that the amplitude can be decomposed as
	\begin{align*}
	\Gamma^{Z' Z \gamma}_{\rho \mu \nu}(r,p,q) = F_1 & \biggl(q^\mu \epsilon[\nu, \rho, p,q]-p^\nu \epsilon[\mu,\rho,p,q]+(p\cdot q)\epsilon[\mu,\nu\,\rho,p]  \\
	&\quad - ((p\cdot q)-M_{Z}^2)\epsilon[\mu,\nu,\rho,q]\biggr)
	+F_2 \, \epsilon[\mu,\nu,\rho,q].
	\end{align*}%
Note that the photon Ward identity is manifest in the above representation of the amplitude.

Contracting with polarization tensors, squaring, averaging over the $Z'$ polarization, and summing over all fermions and final state polarizations, we obtain
\begin{align}
\braket{\left|\mathcal{T}\right|^{2}} =\frac{(M_{Z}^4-M_{Z'}^4)(M_{Z}^2-M_{Z'}^2)}{2 M_{Z}^2 M_{Z'}^2}\left|\sum_f (2 F_{1,f} M_{Z}^2+F_{2¸f})\right|^2,
\end{align}
here $F_{1,f},~F_{2,f}$ denote the form factor contributions from each fermion.

The decay rate is then given by
\begin{align}
\Gamma^{Z' Z \gamma}=\frac{1}{16 \pi M_{Z'}} \left(1 - \left(\frac{M_{Z}}{M_{Z'}}\right)^2\right) \braket{\left|\mathcal{T}\right|^{2}}.
\end{align}

In practice most of the fermion masses can be taken to vanish and only the top-quark mass is assumed to be finite. There is however a subtle issue on how to explicitly perform the massless limit, we refer the reader to the discussion in appendix \ref{app:massless}.

\subsubsection{Forbidden processes}\label{sssec:forbidden}
While the $Z'\rightarrow Z\gamma$ and $Z'\rightarrow Z Z$ processes are allowed and can be observed, a process such as $Z'\rightarrow\gamma \gamma$ is forbidden by the Landau-Yang theorem, and does not contain an anomaly. This can be seen from the only (possibly) non-zero anomaly trace
	\begin{align*}
	\mathrm{Tr}\left(\mathrm{U}(1)_z \mathrm{U}(1)_\text{em}^2\right) &\sim \left(z_\ell-z_e+3\left(\frac{4}{9}(z_q-z_u)+\frac{1}{9}(z_q-z_d)\right)\right) \\
        &\sim \left(z_H+\frac{1}{3}\left(-4 z_H+z_H\right)\right)
	=0.
	\end{align*}%
While the $Z'\rightarrow \gamma^\star \gamma$ process is interesting in its own right, it does not receive contribution from anomalies in this class of models.

\subsection{Branching ratios}\label{ssec:BR}
\begin{figure}[ht]
	\centering
	\captionsetup[subfigure]{}
	\subfloat[][]{\includegraphics[width=0.49\textwidth]{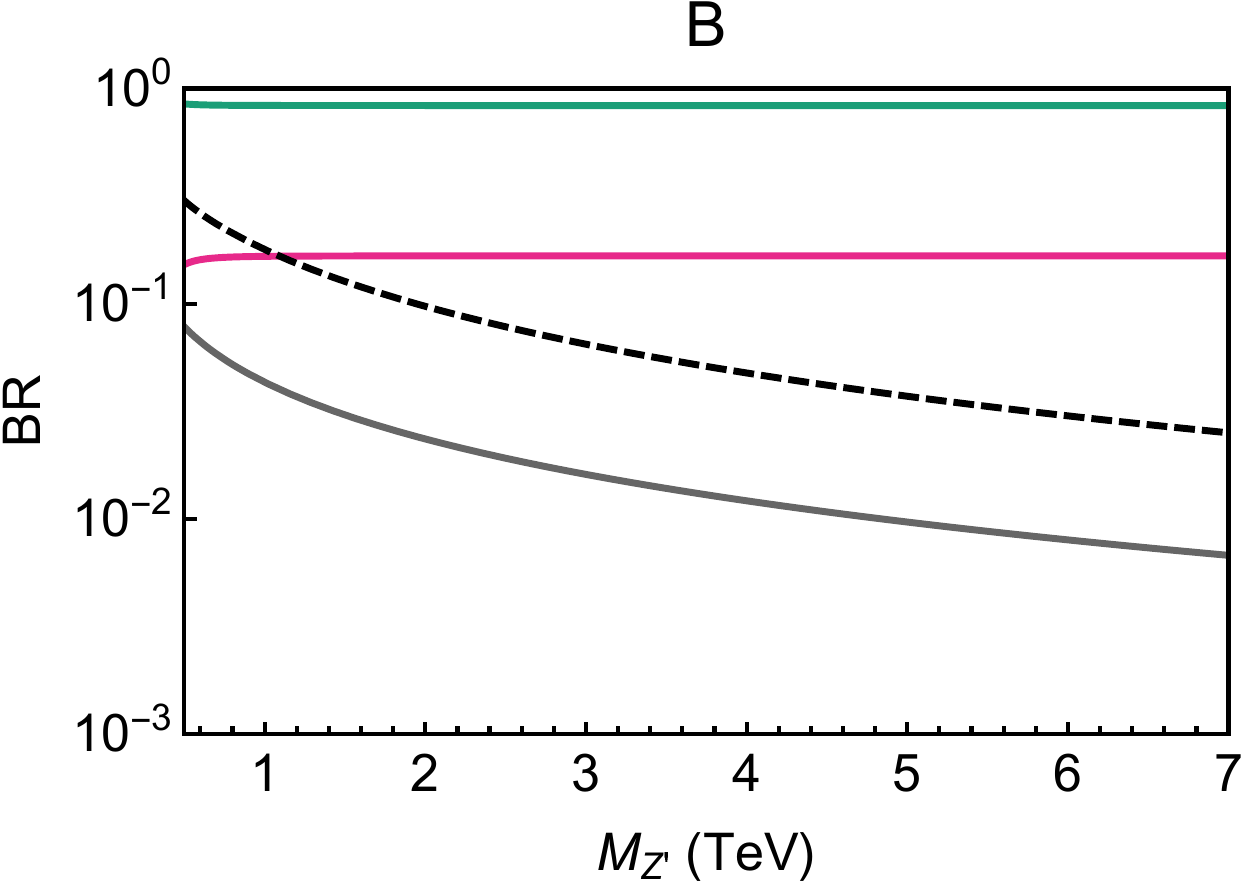}\label{sfig:BRB}}~~
	\subfloat[][]{\includegraphics[width=0.49\textwidth]{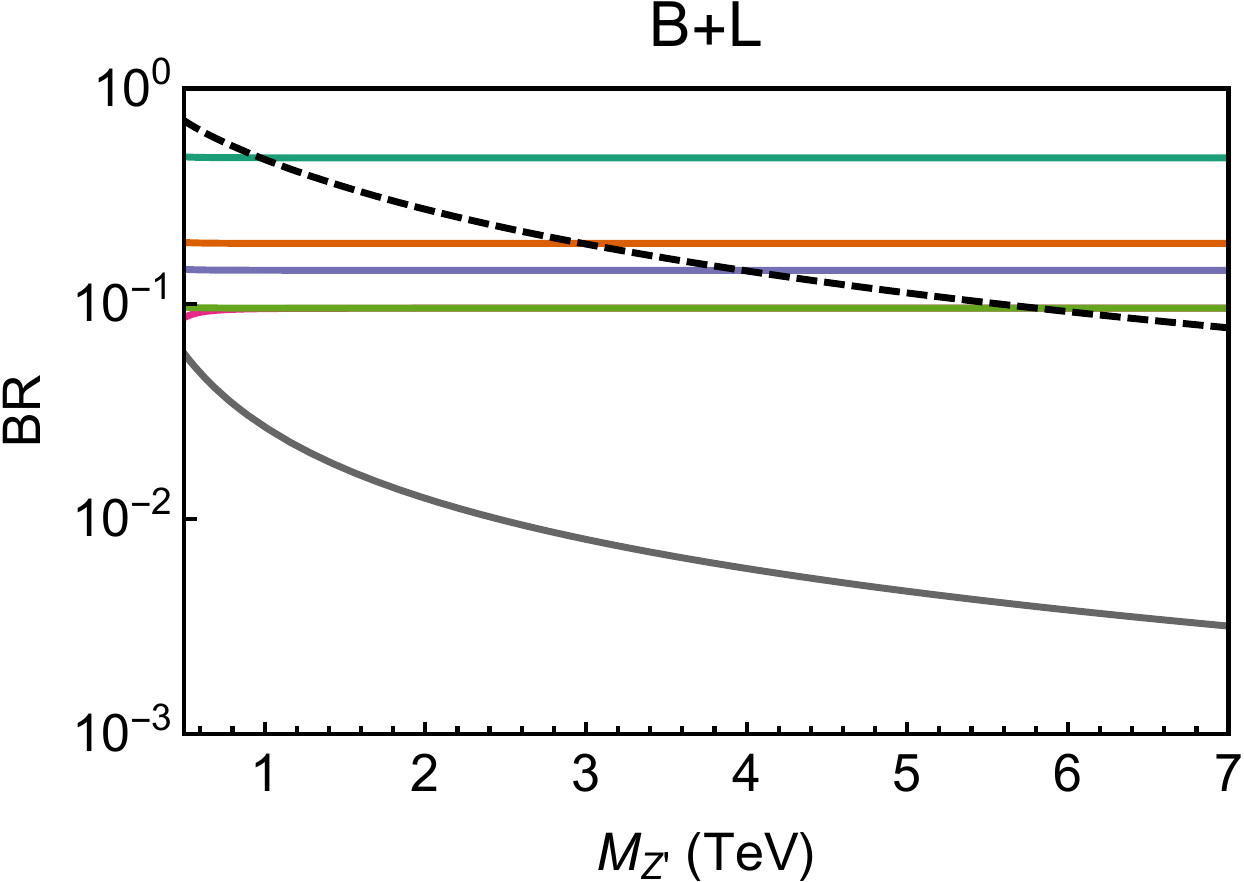}\label{sfig:BRL}}
	\\
	\subfloat[][]{\includegraphics[width=0.49\textwidth]{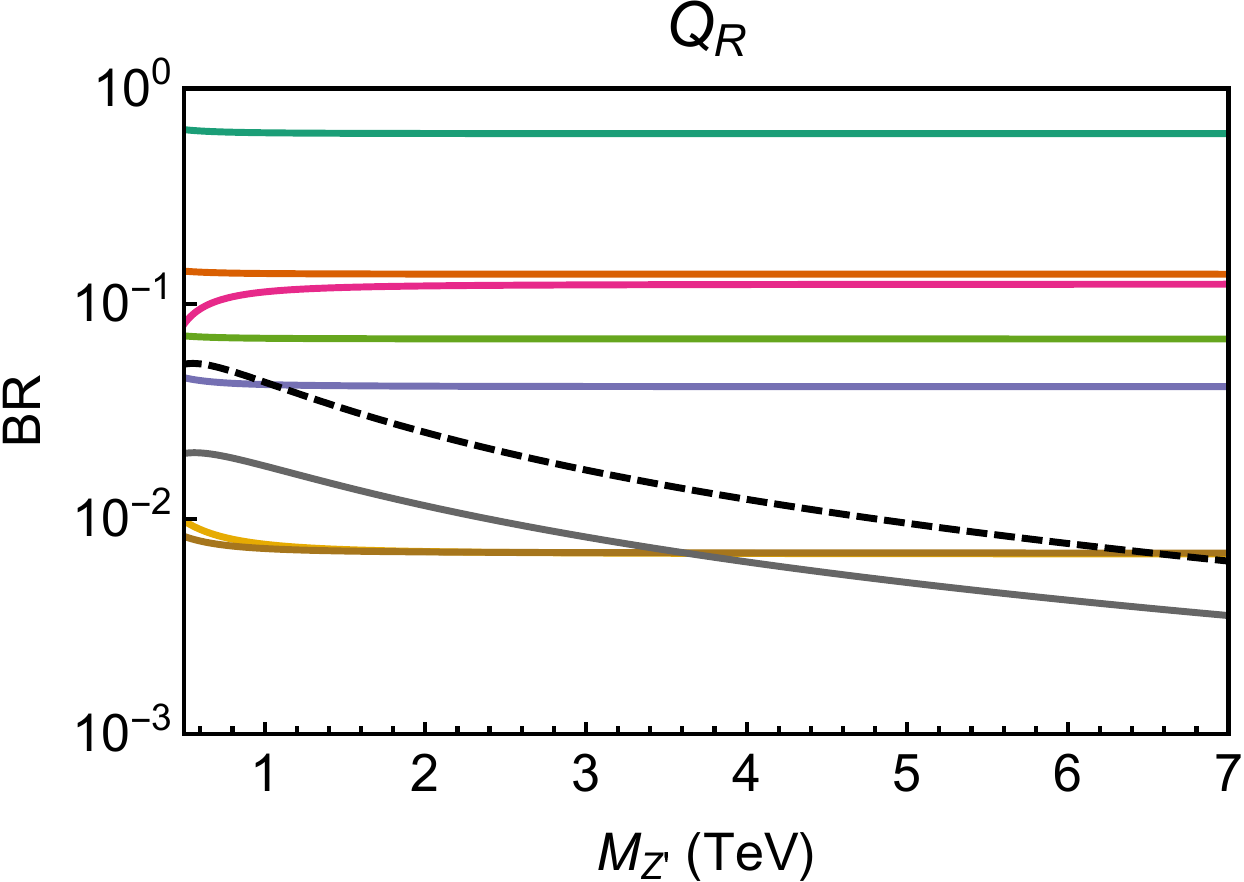}\label{sfig:BRQR}}~~
	\subfloat[][]{\includegraphics[width=0.49\textwidth]{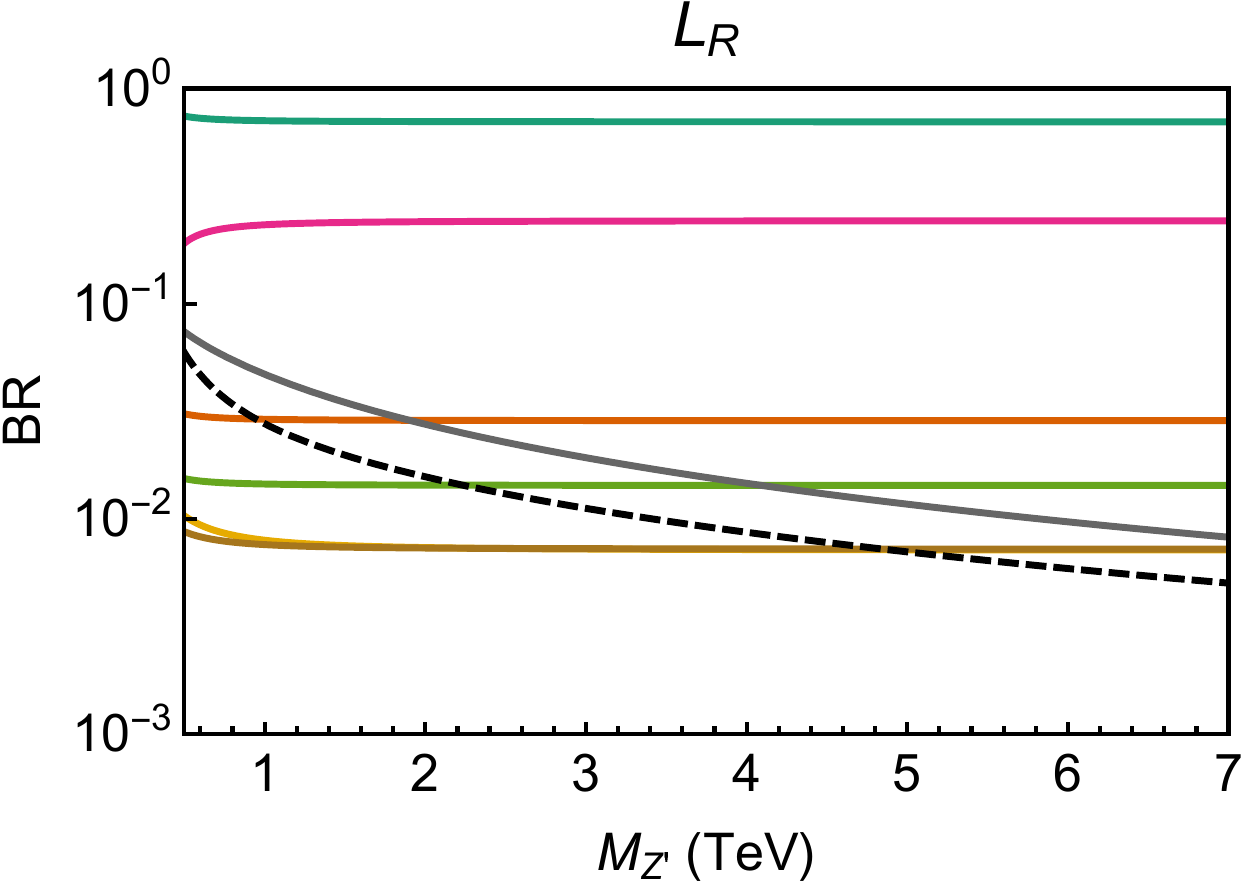}\label{sfig:BRLR}}
	\\
	
	\hbox{\hspace{27ex}\includegraphics[width=0.5\textwidth]{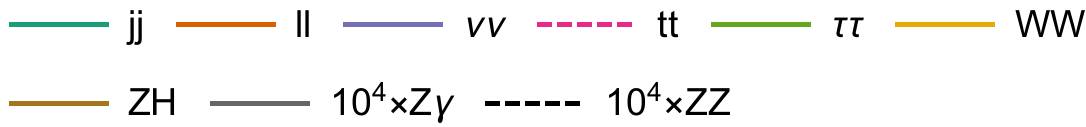}}
	\caption{Branching ratios of $Z'$ as functions of $M_{Z'}$ for the benchmark models given in table\room\ref{tab:benchmodels}. The gauge coupling is chosen to the representative value $ g_{z}=0.2 $. Note the enhancement of a factor of $ 10^{4} $ of the loop-suppressed branching ratios.}
	\label{fig:branchings}
\end{figure}

\begin{table}
	\centering
	\caption{The benchmark models considered in this paper; $ Q_{R} $ refers to right-handed quarks, and $ L_{R} $ to right-handed leptons.}\label{tab:benchmodels}
	\begin{tabular}{c|ccc}
		& $ z_{H} $ & $ z_{q} $ & $ z_{\ell} $\\
		\hline
		$ B $ & $ 0 $ & $ 1/3 $ & $ 0 $\\
		$ B+L $ & $ 0 $ & $ 1/3 $ & $ 1 $\\
		$ Q_{R} $ & $ 1/2 $ & $ 0 $ & $ -1/2 $\\
		$ L_{R} $ & $ 1 $ & $ 1/3 $ & $ 0 $\\
	\end{tabular}
\end{table} 
In Fig.~\ref{fig:branchings}, we show the BRs of $Z'$ as functions of $M_{Z'}$ for the four benchmark models defined in table\room\ref{tab:benchmodels}. The $Z'\rightarrow Z Z$ and $Z'\rightarrow Z \gamma$ branching ratios are multiplied by an extra factor of $ 10^{4} $ for readability. Note that there is no tree-level $Z\leftrightarrow Z'$ mixing in both the $ B $ and the $ B+L $ model -- hence the tree-level decays to $ W^{+}W^{-} $ and $ ZH $ are not present. These decays will be loop-suppressed, presumably on the same order as the $Z'\rightarrow ZZ  $ and $ Z'\rightarrow Z\gamma $ decays, but we have not calculated them since they are not of interest to us.

As can be seen in figure\room\ref{sfig:BRB}, the $Z'$-boson of the $B$-model is leptophobic in nature and therefore dominantly decays to dijets. Hence, the dijet resonance search data is the most important in constraining the $B$-model. 
\section{Collider phenomenology}\label{sec:results}

%%%%%%%%%%%% TM %%%%%%%%%%%%%
\subsection{Exclusion limits}\label{ssec:excl}

At the LHC, a $Z'$ can be produced from $\bar{q}q$ fusion and the production cross section $pp\to Z'$ 
at a fixed collider center-of-mass energy (CME) $\sqrt{s}$ can be parametrized as
\begin{align}\label{eq:cspar}
\sg\lt(M_{Z'},g_z,z_{q},z_{H}\rt)=\frac{g_z^2}{4} \lt[ a^u\lt(M_{Z'}\rt)\lt\{z_{q}^2+\left(z_{q}+z_{H}\rt)^{2}\rt\} +
a^d\lt(M_{Z'}\rt)\lt\{z_{q}^2+\left(z_{q}-z_{H}\rt)^{2}\rt\}\rt],
\end{align}
where the mass-dependent functions (also dependent on $\sqrt{s}$) $a^u$ and $a^d$ include contributions from all the 
up-type ($u,c$) and down-type ($d,s,b$) quarks in the proton, respectively. Another free parameter in our set-up 
is $z_\ell$ which would not appear in the production cross section. Notice that although $Z'$ couples differently to
the left-handed and right-handed components of a quark, the functions $a^u\lt(M_{Z'}\rt)$ and $a^d\lt(M_{Z'}\rt)$ do not depend on the chirality. 
To obtain these functions (numerically), we interpolate the production cross sections of $Z'$ computed for different $M_{Z'}$
for a reference $Z'\bar{q}q$ coupling. We use the NN23LO~\citep{Ball:2012cx} PDF set to compute $\sg(pp\to Z')$ at leading 
order (LO) at a fixed factorization ($\mu_F$) and renormalization ($\mu_R$) scale $\mu_F=\mu_R=M_{Z'}$. We perform this
calculation using the \textsc{MadGraph5}~\cite{Alwall:2014hca} event generator, where the model files are generated
using \textsc{FeynRules}~\cite{Alloul:2013bka}. The calculation of the relevant BRs is discussed in subsection~\ref{ssec:BR} and we assume the 
narrow width approximation (NWA) is valid to factorize $\sg(pp\to Z'\to XY)$ into $\sg(pp\to Z')\times BR(Z'\to XY)$. For more
accurate exclusion, capturing higher-order effects, we multiply the LO $\sg(pp\to Z')$ by a constant next-to-leading order 
(NLO) QCD K-factor of 1.3 for any $M_{Z'}$~\cite{Gumus:2006mxa}. In our analysis, we consider the new gauge coupling $g_z$ as
a free parameter, and for large values of $g_z$ (or for large values of various effective couplings) electroweak corrections 
might be important in addition to the QCD corrections. Considering those higher-order effects is beyond the scope of the present paper.

We use results from the two direct $Z'$ resonance searches in the dilepton and dijet channels at the 13 TeV LHC. In order 
to set exclusion limits on $Z'$ parameters, we compare the 95\% confidence level (CL) upper limits on the $\sg\times BR$
of $Z'$ set by the ATLAS and CMS collaborations in these two channels with our model predictions. Here we use ATLAS 
dilepton~\cite{Aaboud:2017buh} and dijet~\cite{Aaboud:2017yvp} data and CMS dijet data~\cite{CMS:2017xrr}, both available for $\sim 36$ fb$^{-1}$ integrated luminosity.

In addition to the collider data, we also use tree-level $T$-parameter constraints (as discussed in~\cite{Appelquist:2002mw}) for exclusion limits. The current constraint on the $T$-parameter is $0.08\pm 0.12$~\cite{Olive:2016xmw} which has been used in
our analysis. Another constraint on $Z'$ models might come from the $Z$-boson width measurements, however the $Z$-boson width constraints are quite similar to the $T$-parameter. Therefore, we have
ignored the $Z$-boson width constraint in this paper. We expect any other electroweak precision constraints to be subdominant, since they all enter at the one loop level.

\begin{figure}[ht]
\centering
\captionsetup[subfigure]{}
\subfloat[][]{\includegraphics[height=6.55cm,width=6cm]{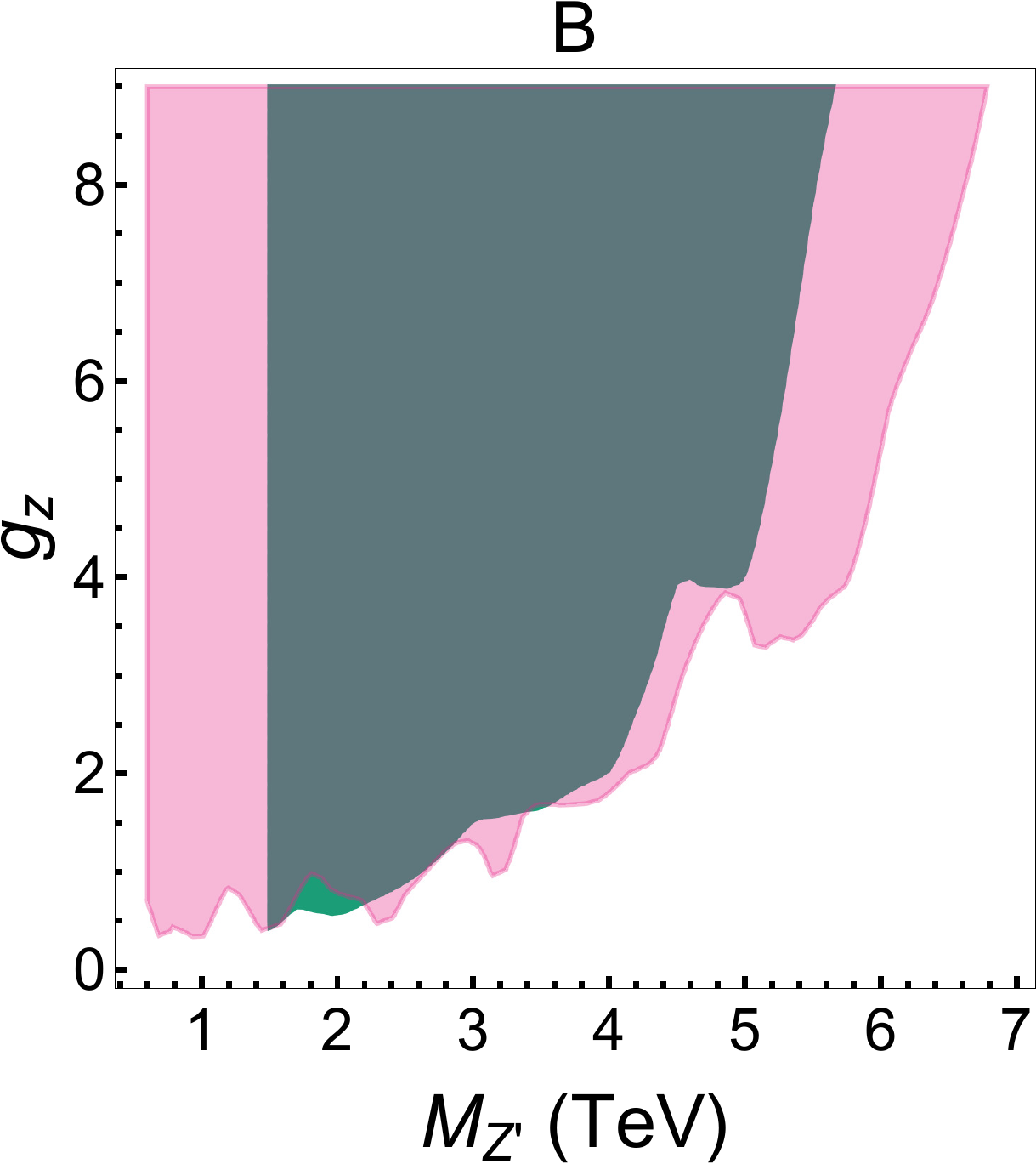}\label{sfig:B}}\hspace{1cm}
\subfloat[][]{\includegraphics[height=6.55cm,width=6cm]{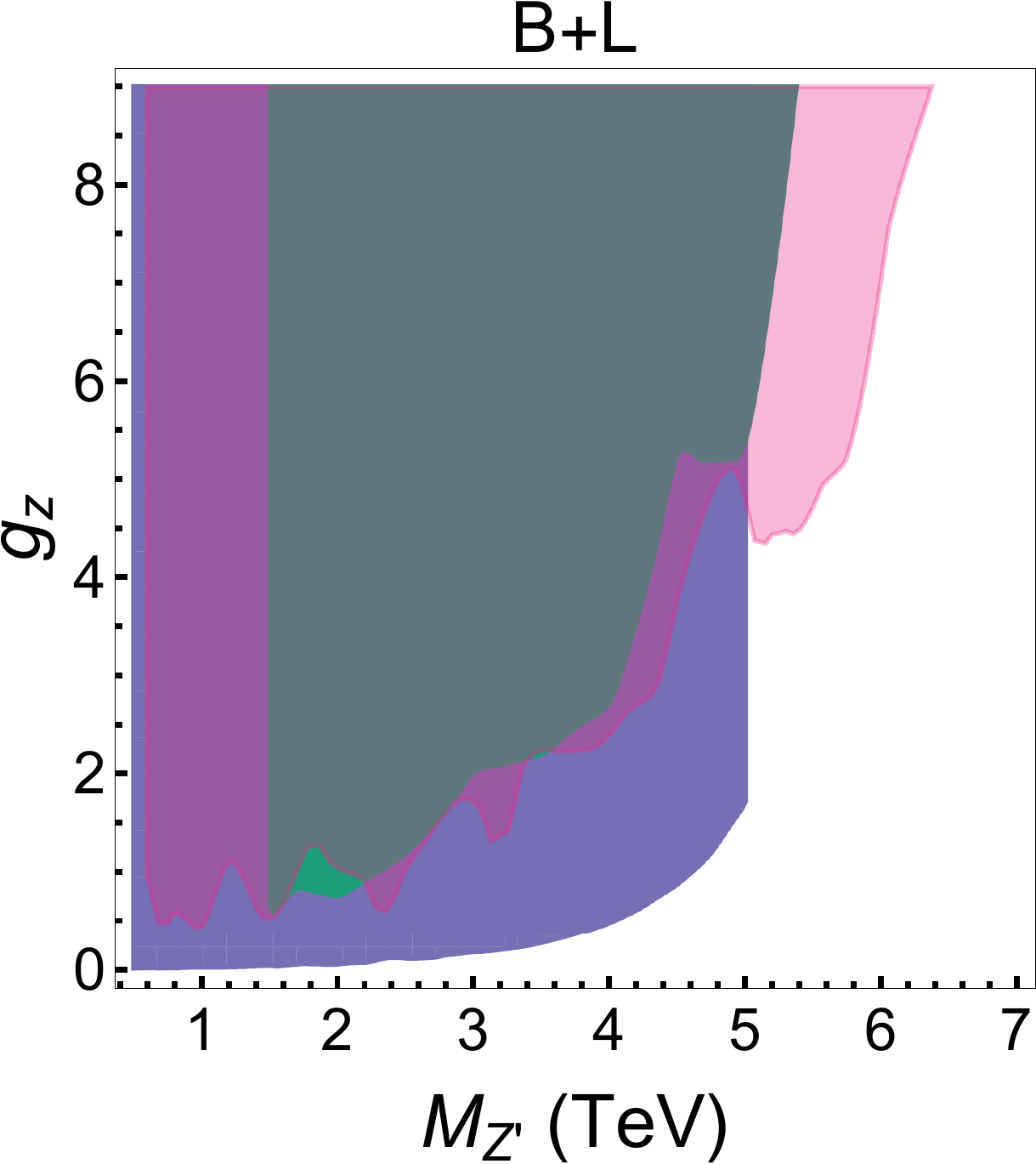}\label{sfig:BpL}}\\
\subfloat[][]{\includegraphics[height=6.55cm,width=6cm]{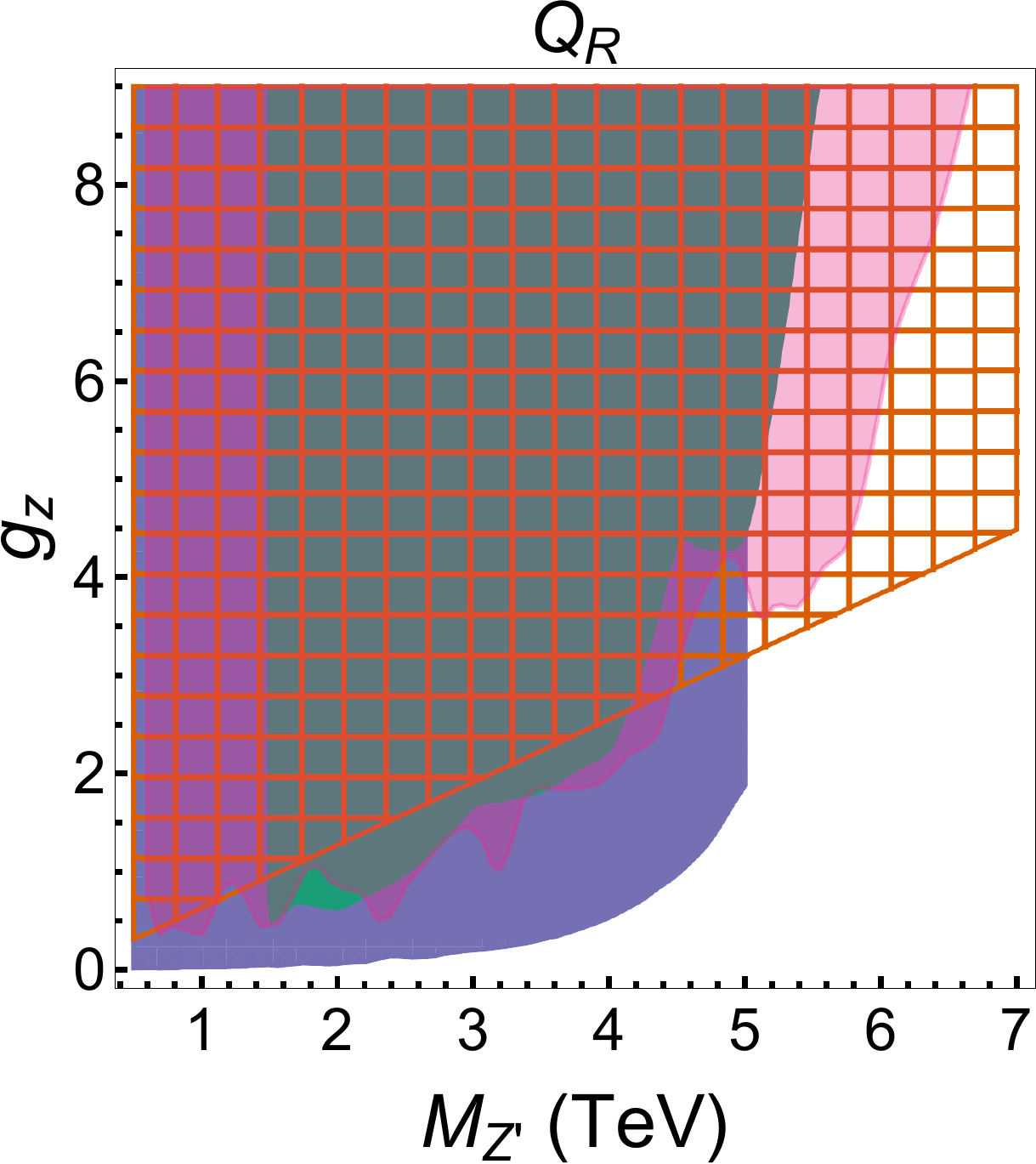}\label{sfig:QR}}\hspace{1cm}
\subfloat[][]{\includegraphics[height=6.55cm,width=6cm]{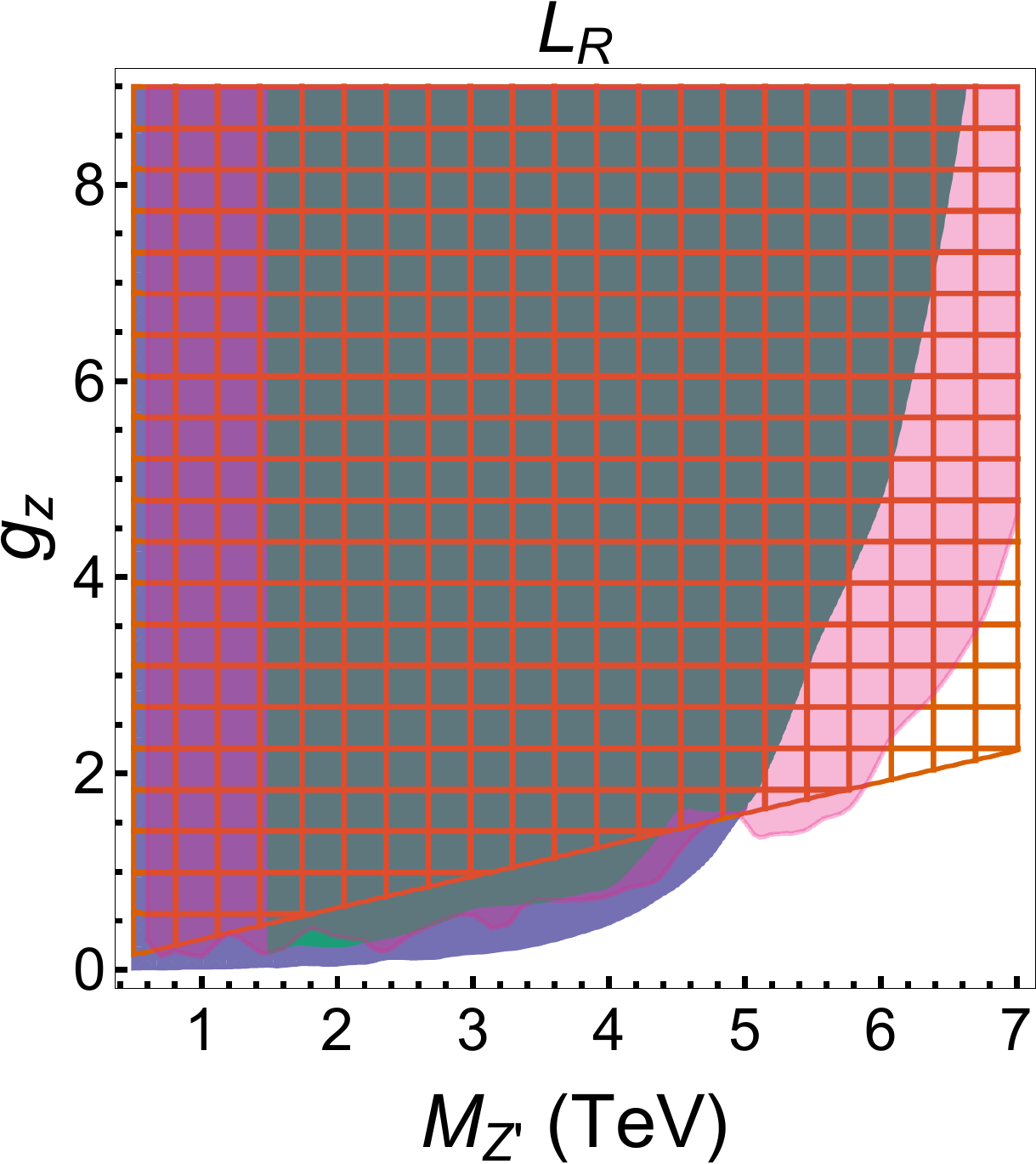}\label{sfig:LR}}\\	
\includegraphics[width=0.8\textwidth]{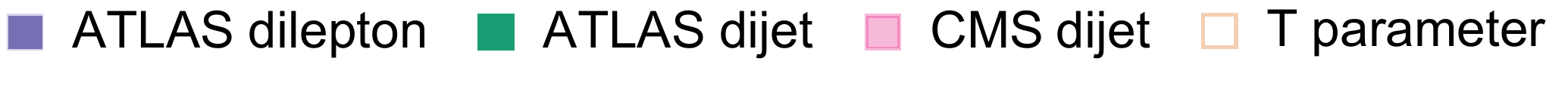}
\caption{The marked regions are excluded from various experimental constraints. The lilac filled region corresponds to 
$\mc{R}>1$, where $\mc{R}=(\sg\times BR_{ll})^\text{th}/(\sg\times BR_{ll})^\text{obs}_\text{ATLAS}$ and 
$(\sg\times BR_{ll})^\text{th}$ and $(\sg\times BR_{ll})^\text{obs}_\text{ATLAS}$ denote our prediction and the observed
95\% CL upper limit set by ATLAS using dilepton resonance search data at the 13 TeV LHC~\cite{Aaboud:2017buh}, respectively. The 
filled teal region is the similar comparison with the 13 TeV ATLAS dijet data~\cite{Aaboud:2017yvp}; the transparent magenta region is the corresponding comparison with the 13 TeV CMS dijet data~\cite{CMS:2017xrr}. The orange/red grid-covered
region is excluded from the $T$-parameter constraints.}
\label{fig:excl}
\end{figure}

In figure~\ref{fig:excl}, we show sample exclusion plots in the $M_{Z'}-g_z$ plane for four selected $Z'$ models discussed 
in section~\ref{sec:models}. As an illustration, we pick the $B$ and $B+L$ models from the $z_H=0$ category and the $Q_R$ and $L_R$ models from the $z_H\neq 0$ category. The tree-level $Z\leftrightarrow Z'$ mixing does not arise in models with $z_H=0$ ($Z\leftrightarrow Z'$ mixing can still arise in these models at loop level). Therefore, the tree-level $T$-parameter constraint is not applicable for
this category. In the $B$ model (figure~\ref{sfig:B}), the $Z'$ couples only to quarks and hence dilepton data is not relevant to
constrain this model. We observe that dilepton data, wherever applicable, can constrain various $Z'$ models severely. For the $L_R$
model (figure~\ref{sfig:LR}), dijet data is also very effective in constraining the model.

\subsection{Interesting signatures}\label{ssec:sig}

As discussed before, possible signatures of a GS $Z'$ can be seen in the $ZZ$ and $Z\gm$ decay modes. 
These decay modes, however, have tiny BRs because the leading contribution from the GS terms is at the one-loop level.
We have already 
seen in subsection~\ref{ssec:BR} that the BRs of $Z'$ to $ZZ$ and $Z\gm$ modes are tiny in comparison with the dilepton 
and dijet decay modes and, therefore, observing these
modes at the LHC could be very challenging. Both the BRs (see section\room\ref{sec:BR}) for the $ZZ$ and the $Z\gamma$ 
modes and the production cross section of $Z'$ decrease for large $M_{Z'}$. Hence, the small mass region offers the best chance of observing these decays, On the other hand the collider and EW precision bounds discussed in subsection~\ref{ssec:excl} are quite constraining in the small mass region and it can be hard to find the best parameter points. 

In order to find the optimal region of the parameter space, we perform a random scan over $M_{Z'},g_z,z_H,z_q$ and $z_\ell$
in the ranges
\begin{align}
\label{eq:scanRange}
0.5~\mathrm{TeV} < M_{Z'} < 0.8~\mathrm{TeV},~~0.01 <g_z<0.3,~~0 <z_H,z_\ell <1.0,~~0 < z_q < 8, 
\end{align}
which is where most of the allowed points lie.  

\begin{figure}[phtb]
\centering
\captionsetup[subfigure]{}
\subfloat[][]{\includegraphics[width=0.49\columnwidth]{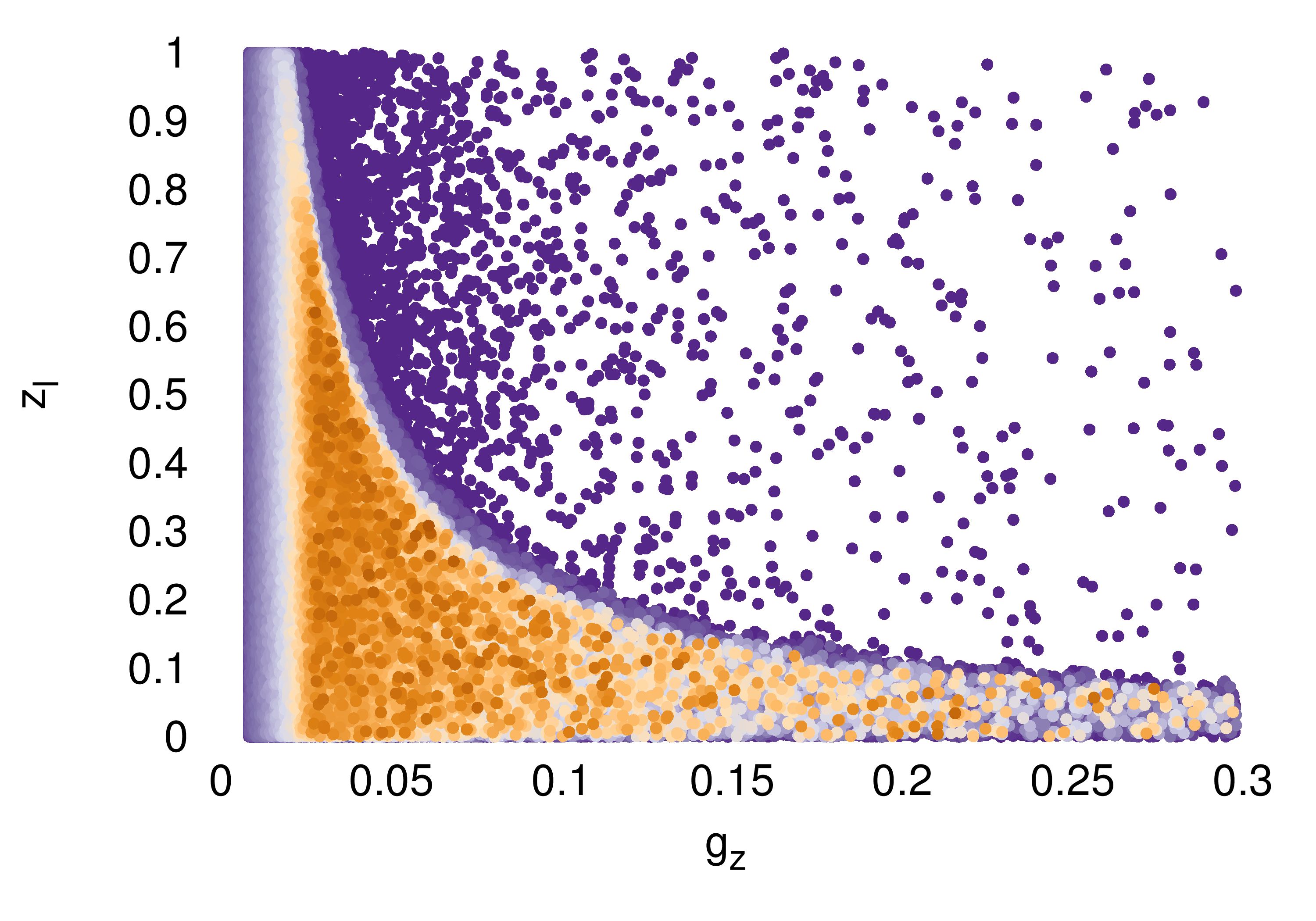}\label{sfig:ZZ-zlgz}}
\subfloat[][]{\includegraphics[width=0.49\columnwidth]{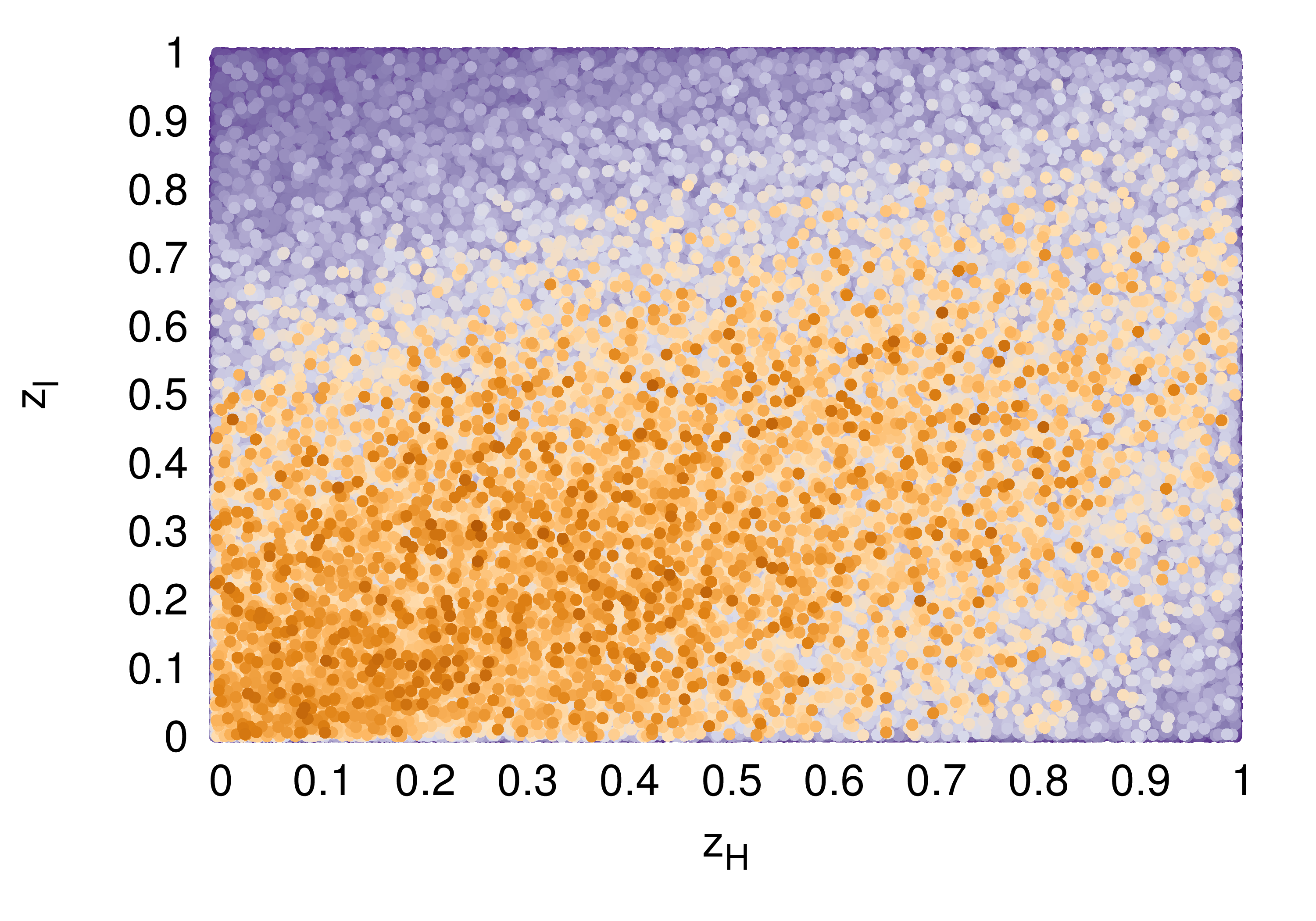}\label{sfig:ZZ-zlzh}}\\
\subfloat[][]{\includegraphics[width=0.49\columnwidth]{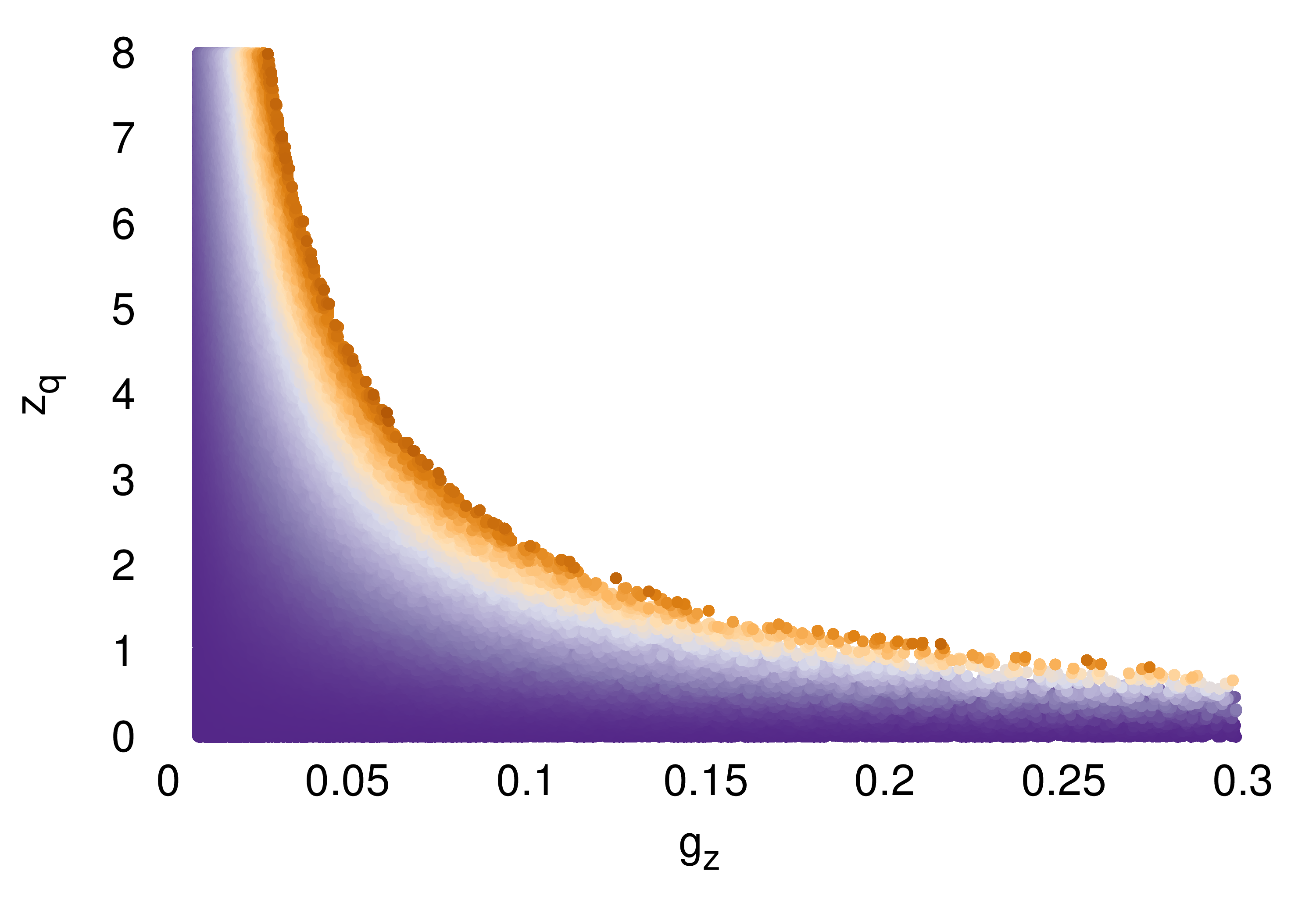}\label{sfig:ZZ-zqgz}}	
\subfloat[][]{\includegraphics[width=0.49\columnwidth]{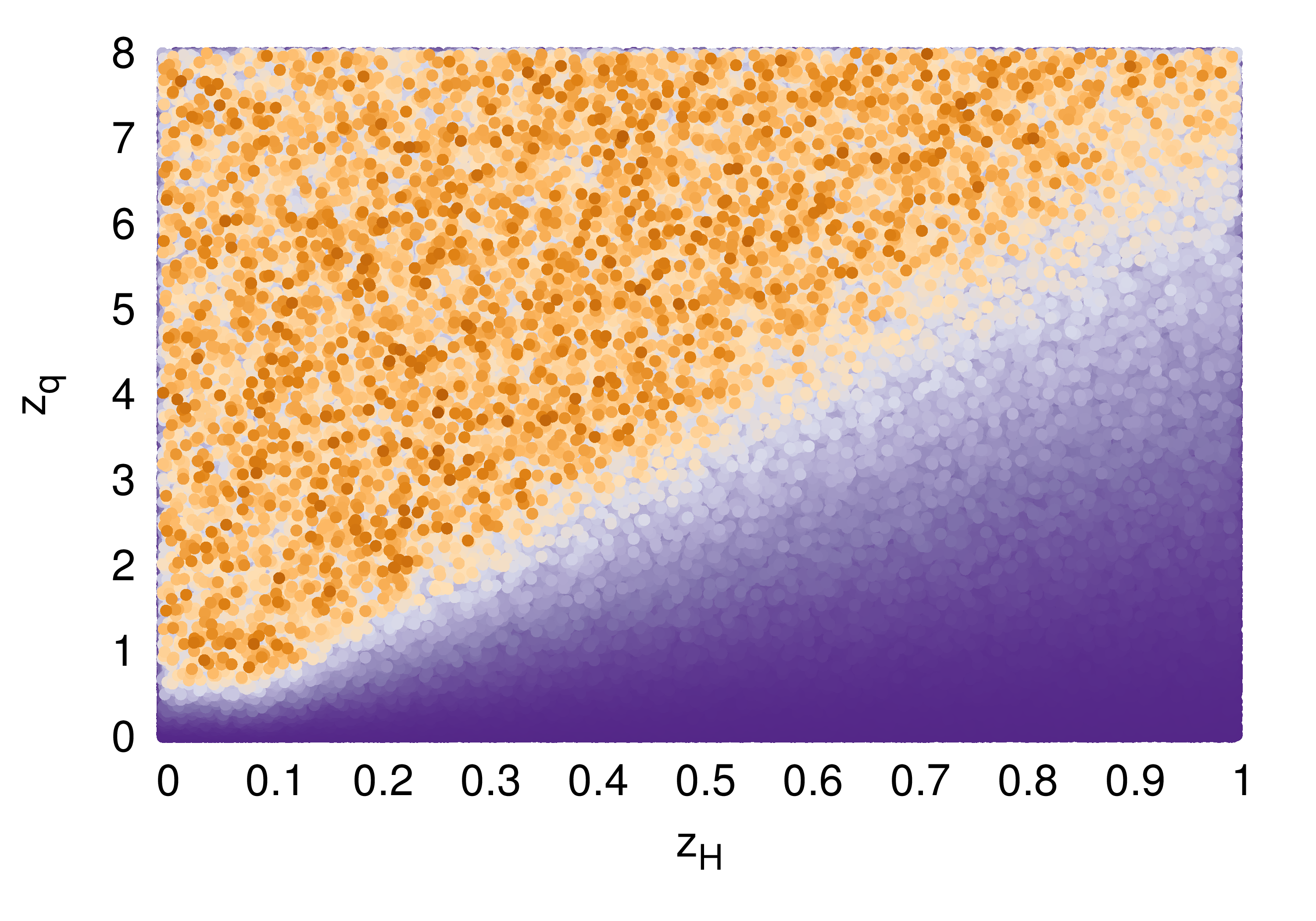}\label{sfig:ZZ-zqzh}}\\
\subfloat[][]{\includegraphics[width=0.49\columnwidth]{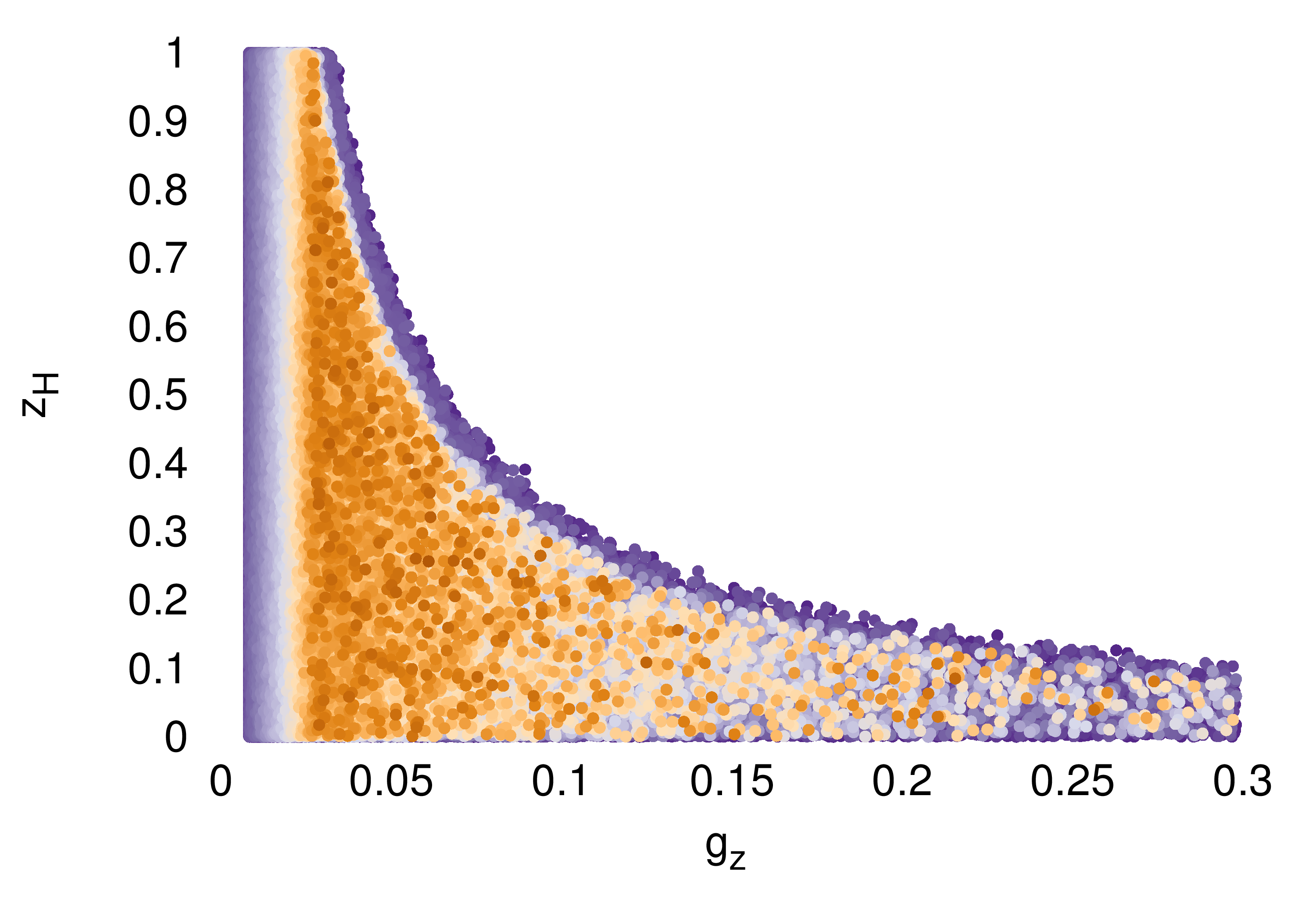}\label{sfig:ZZ-zhgz}}
\subfloat[][]{\includegraphics[width=0.49\columnwidth]{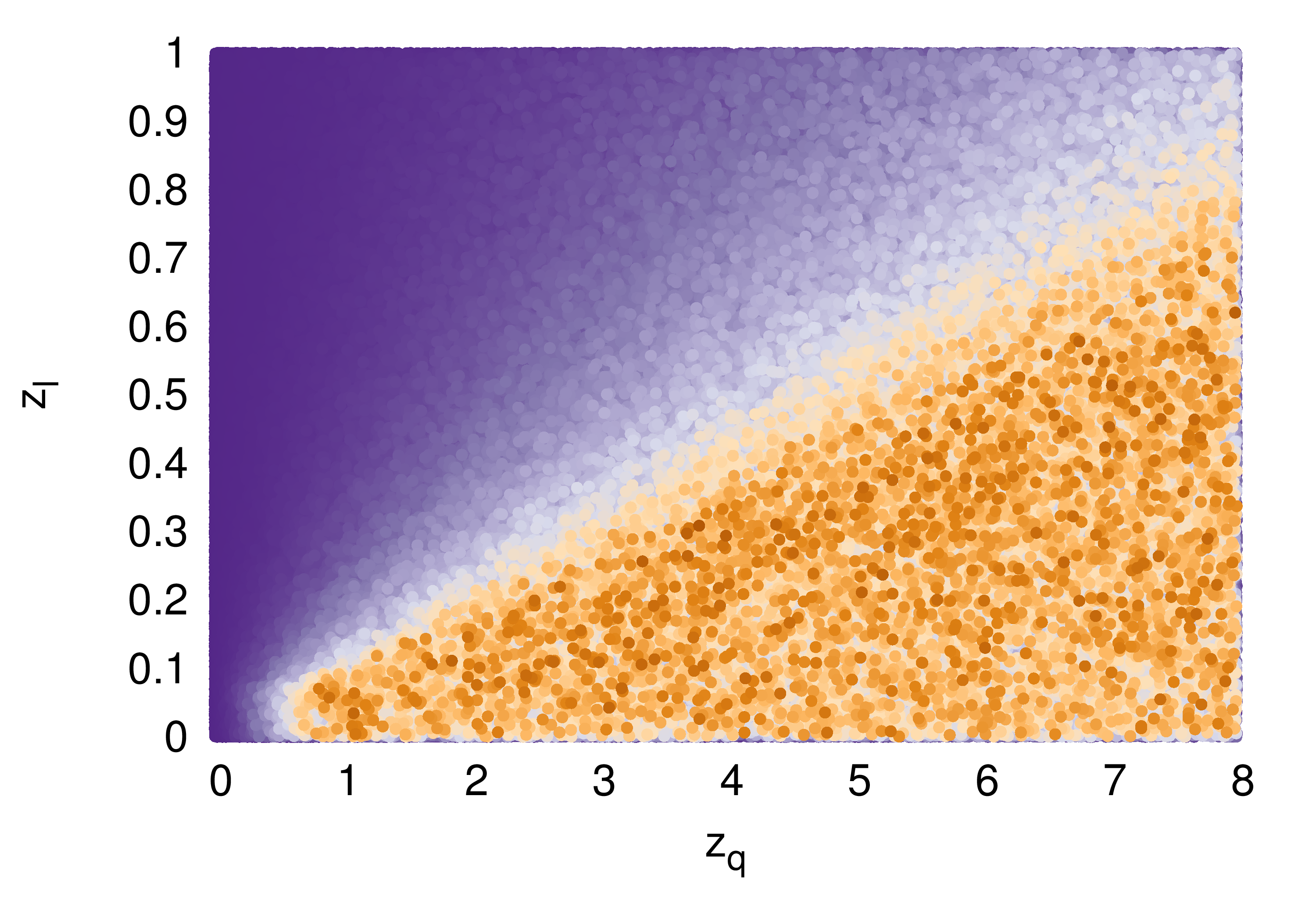}\label{sfig:ZZ-zqzl}}\\
\includegraphics[width=0.8\columnwidth]{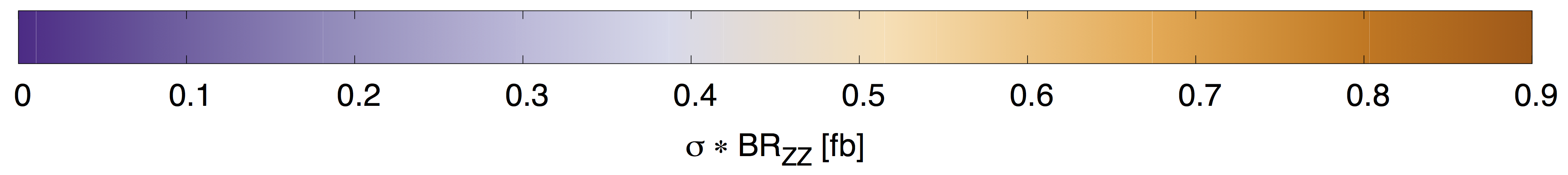}
\caption{2D heat maps of $\sg(pp\to Z')\times BR(Z'\to ZZ)$ in fb against different combinations of the parameters $g_{z},z_H,z_q$ and $z_\ell$ obtained from a random scan of the 
5-D volume defined in Eq.~\eqref{eq:scanRange} and with points shown satisfying the experimental constraints considered in subsection~\ref{ssec:excl}.}
\label{fig:scanHeatZZ}
\end{figure}

In figure~\ref{fig:scanHeatZZ}, we display two-dimensional "heat maps" of $\sg(pp\to Z')\times BR(Z'\to ZZ)$ for different combinations of the parameters. First of all, since BRs of $Z'$ to $ZZ$ and $Z\gm$ are very small, 
$M_{Z'}$ should not be too heavy in order to have sufficiently big $\sg\times BR$ to be observable at the LHC. From figure~\ref{sfig:ZZ-zqzh}, it can be seen that $z_q$ has to be larger than $\sim 0.5$ in order to get a decent cross section -- this bound arises from our upper bound of $g_z$ which ensures that small $z_q$ values limit the production cross section. Additionally, it is noted that in order to have a sizable cross section for larger $z_\ell$ values it is necessary to increase $z_q$, due to the lepton data being more constraining than the dijet data. As a result, $z_q$ has to increase in order to ensure that the dilepton branching stays sufficiently small.

Large $z_q$ values are mainly accompanied by small $g_z$ values in order to keep the production cross section small enough to evade the dijet bounds. Figure\room\ref{sfig:ZZ-zlzh} shows that the most favorable region occurs around the line $z_\ell\sim z_H$. This is because for a given $z_\ell$, the lepton decay width obtains its minimal value for $z_H$ of the order $z_H\sim z_\ell$. 

Note that similar plots can also be made for the $Z\gm$ decay mode. We found that those are identical in structure but with a different scaling. Altogether, the largest cross sections for the $ZZ$ and the $Z\gamma$ channels are around $\sim 0.9$ fb for the former and $\sim 0.25$~fb for the latter around $M_{Z'}\sim 0.5$ TeV. Since the most optimistic cross sections 
for these channels are of the order of $\sim 0.1-1$~fb, it is very hard to detect these decay modes until the high-luminosity LHC. For the HL-LHC we expect a maximum of 3000 fb$^{-1}$ integrated luminosity. Therefore, in the best case scenario, one would expect of the order of $\sim 3000$ $ZZ$ events and $\sim 600$ $Z\gamma$ events that come from GS $Z'$ decay. 
The prospect study of GS $Z'$ in these channels is beyond the scope of this paper.
It should be noted
that the dijet and dilepton bounds can be expected to be significantly improved, if the $Z'$ is not discovered in those channels, for an integrated luminosity of 3000 fb$^{-1}$, which means that the parameter space would shrink significantly.

\subsection{Lepton colliders}\label{subsec:lepcol}

In addition to the high luminosity LHC, 
it is also important to analyze the 
prospects of observing an anomalous $Z'$, either directly or indirectly, in the
context of future lepton colliders such as ILC, FCC-ee, CLIC, etc. This has 
previously been analyzed in the literature on many occasions~\cite{DelAguila:1993rw,Godfrey:1994qk,DelAguila:1995fa,Godfrey:2005pm,Osland:2009dp,Battaglia:2012ez,Han:2013mra}.
In this subsection, we study the reach of the ILC when probing a $Z'$ parameter space, in a model independent way.   

\begin{figure}[H]
\centering
\captionsetup[subfigure]{}
\subfloat[][]{\includegraphics[width=0.33\textwidth]{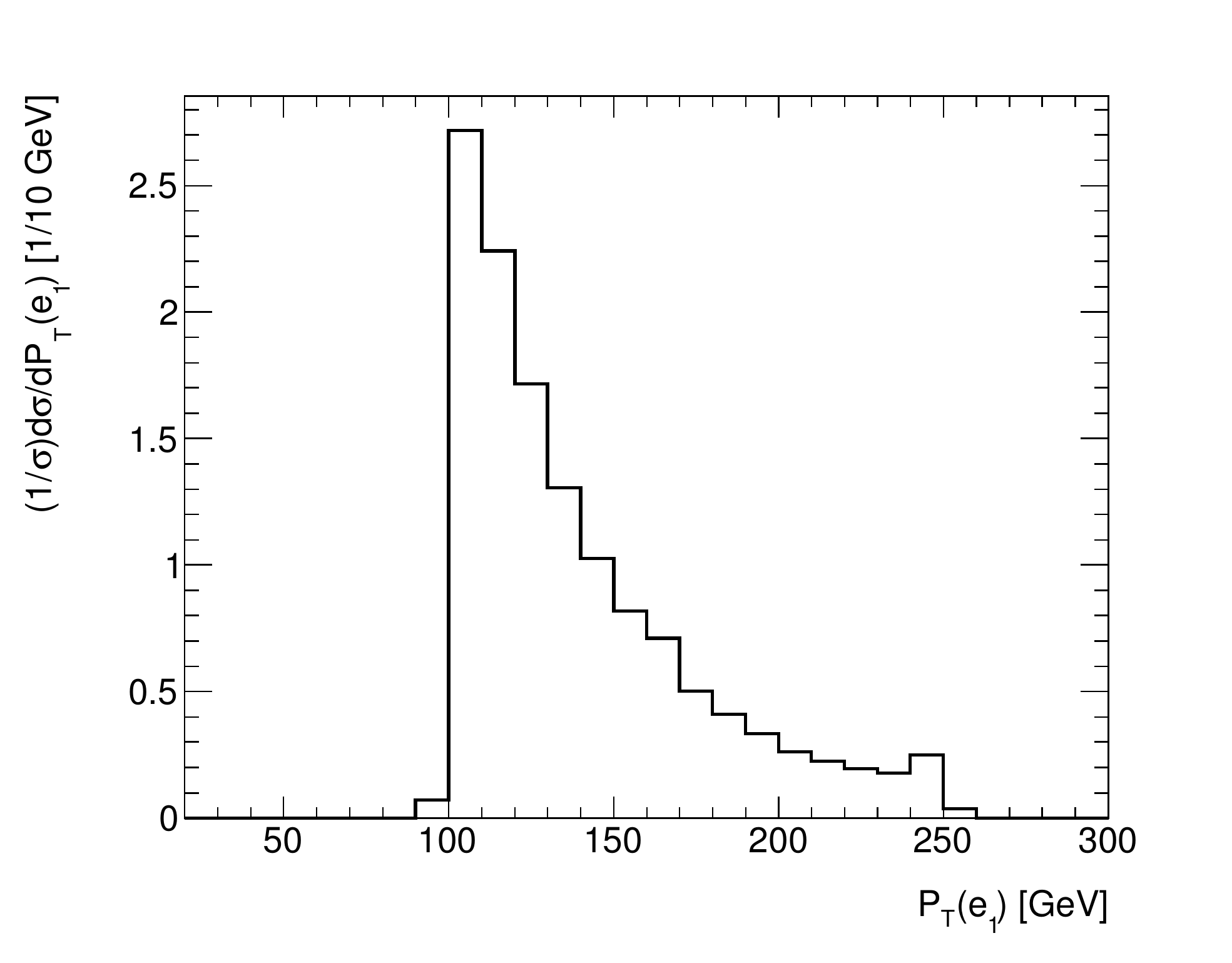}\label{sfig:pTlSM}}
\subfloat[][]{\includegraphics[width=0.33\textwidth]{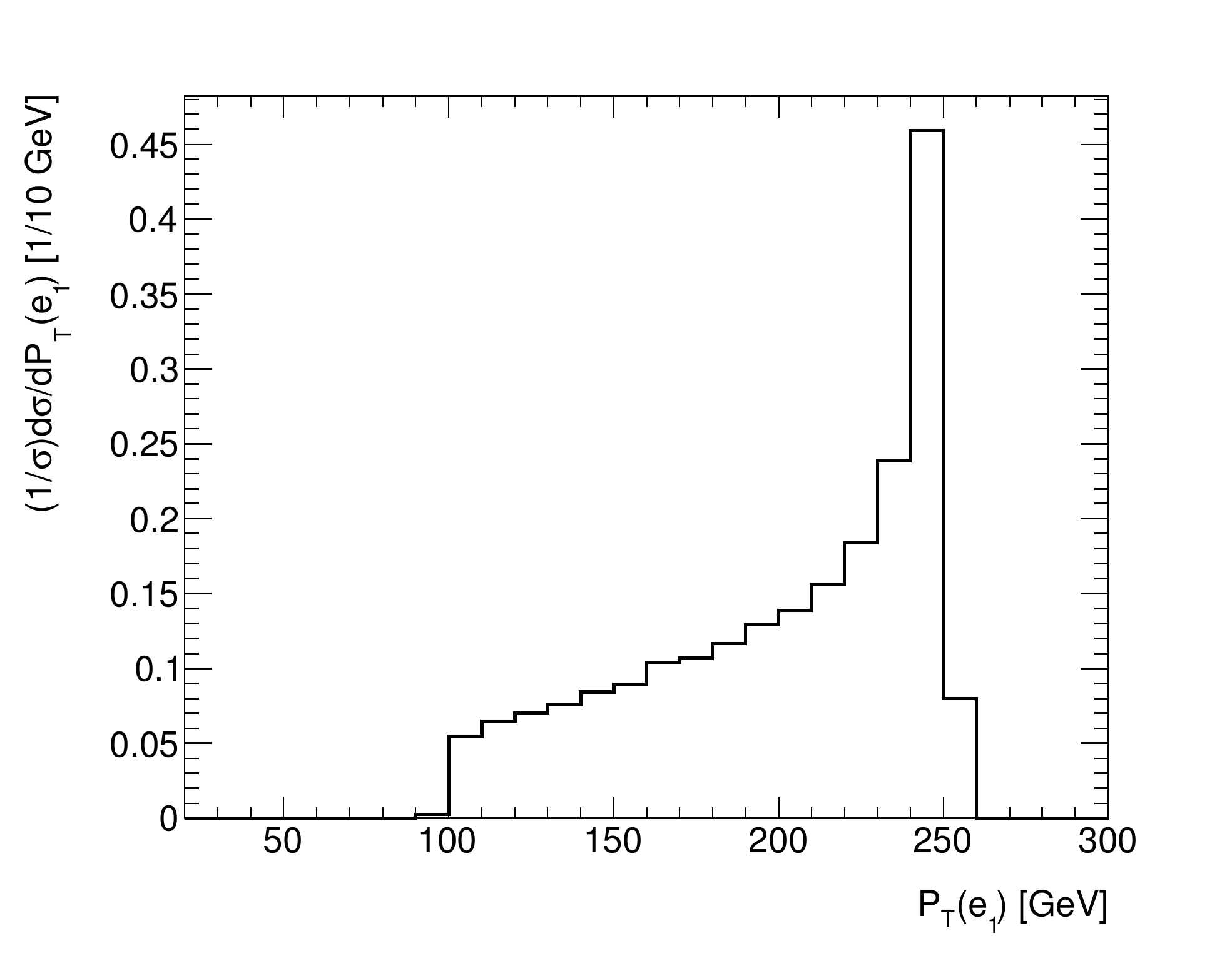}\label{sfig:pTlBSM}}
\subfloat[][]{\includegraphics[width=0.33\textwidth]{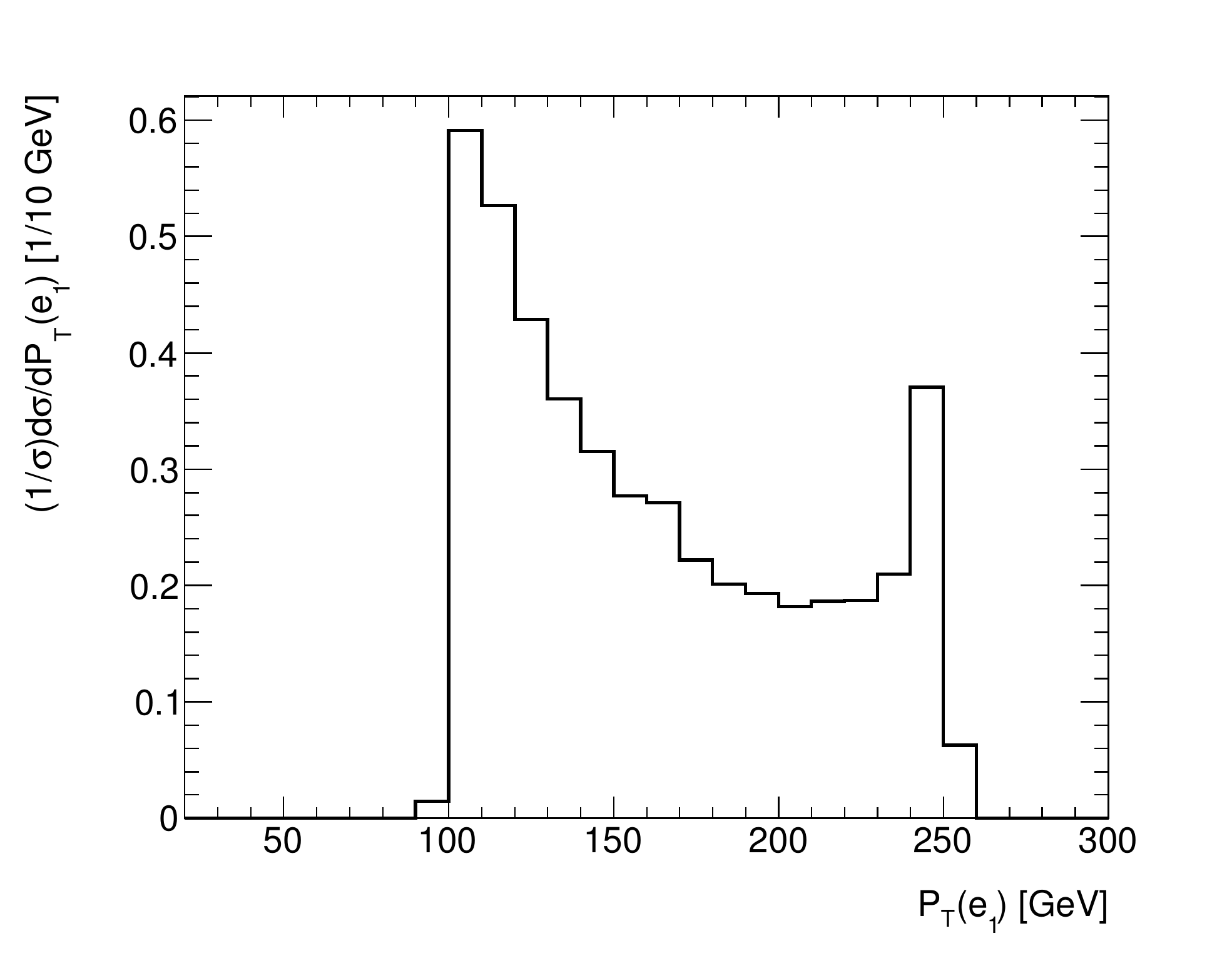}\label{sfig:pTltot}}	
\caption{$p_T$ distribution of the hardest electron for (a) the SM background, (b) only $t$-channel $Z'$
exchange and (c) total $Z'$ and the SM including interference after selecting events by applying
the cut $p_T(e_1),p_T(e_2)>100$ GeV at the 0.5 TeV ILC for $M_{Z'}=1$ TeV.}
\label{fig:pTl}
\end{figure}

In an $e^+e^-$ collider, a $Z'$ can be produced through the $s$-channel as well as 
the $t$-channel exchange. If the collider CME is smaller than the $Z'$ mass,
it is not possible to produce the $Z'$ resonantly on-shell. Therefore, 
the only possible way to observe the hint of $Z'$ signal is indirectly through the interference effects. 
The proposed initial ILC CME is $\sqrt{s}=0.5$ TeV which 
is smaller than the $M_{Z'}$ range of our interest and therefore, a $Z'$ cannot be produced resonantly at the ILC.
It turns out that the reach of various $Z'$ models is better for the 500 GeV ILC than the 14 TeV
LHC~\cite{Osland:2009dp,Han:2013mra}, thanks to the sufficiently large interference 
which does not fall off rapidly with the increase of $M_{Z'}$.
For the model independent analysis, we have, effectively, two free parameters, $M_{Z'}$ and $\kp$
(where $\kp$ is the (total) $Z'e^+e^-$ coupling). In the signal definition,
we include the interference term (which is actually the dominant one and goes as $\kp^2$) 
in addition to the pure new physics term that varies as $\kp^4$. Generally, 
the new physics coupling $\kp$ is expected to be small (less than unity) and hence the interference 
term actually dominates in the signal. The total $e^+e^-\to e^+e^-$ (including the
SM and BSM parts) cross section can be expressed as
\begin{figure}
\centering
\includegraphics[height=7cm,width=8cm]{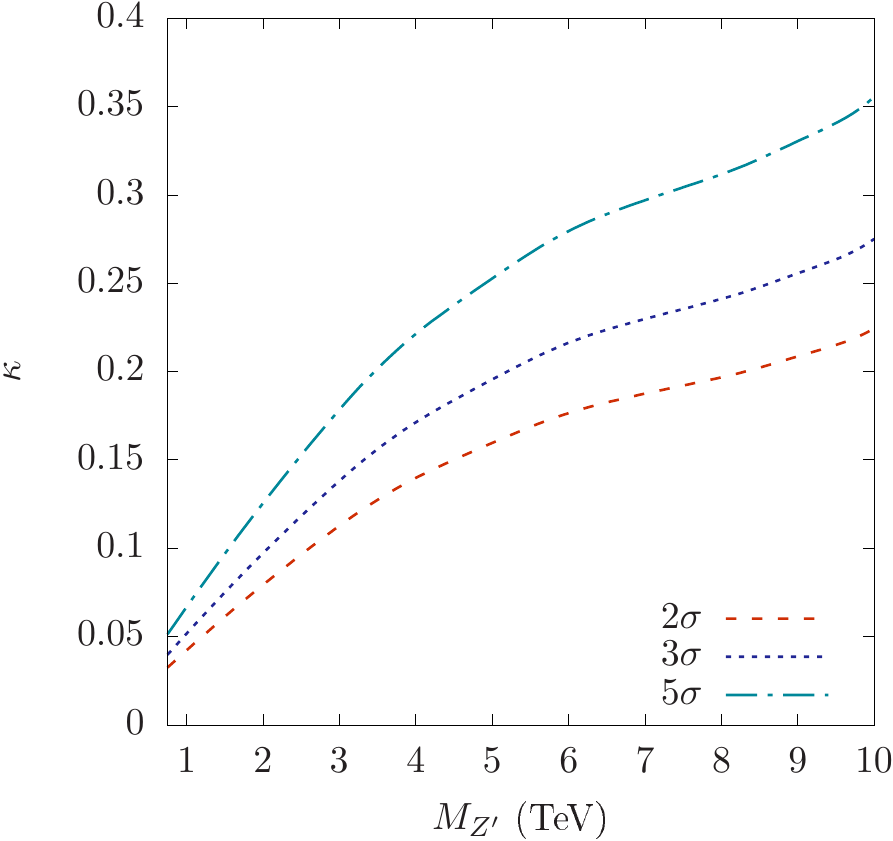}
\caption{Confidence level contours with significance $2\sg$, $3\sg$ and $5\sg$ in the $M_{Z'}-\kp$ plane at the 0.5 TeV ILC with 100 fb$^{-1}$ integrated luminosity.}
\label{fig:significance}
\end{figure}
\begin{align}
\sg_{tot}(e^+e^-\to e^+e^-) = \sg_{SM} + \kp^2\sg_{I}(M_{Z'}) 
+ \kp^4\sg_{BSM}(M_{Z'})
\end{align}
where $\sg_B = \sg_{SM}$ acts as the SM background and the signal is defined as
$\sg_S = \kp^2\sg_{I} + \kp^4\sg_{BSM}$. The dominant SM background comes
from the $s$- and $t$-channel photon and $Z$ exchange processes. In case for 
signal where a massive $Z'$ is exchanged, the $p_T$ distributions of the outgoing
electrons peak around $\sqrt{s}/2$ whereas for the background it peaks towards the
lower side of $p_T$. In figure~\ref{fig:pTl}, we show the $p_T$ distributions of the hardest electron after applying
a strong preselection cut of $p_T(e_1),p_T(e_2)>100$ GeV for $M_{Z'}=1$ TeV at the 0.5 TeV
ILC (in addition, we also apply a few basic cuts \emph{viz.} $|\eta(e_1)|,|\eta(e_2)| < 2.5$, $\Delta R(e_1,e_2)>0.4$). In figures~\ref{sfig:pTlSM} and \ref{sfig:pTlBSM}, we show distributions for the SM background and for the
pure BSM part, respectively. While in figure~\ref{sfig:pTltot}, we show the same for the total 
$e^+e^-\to e^+e^-$ process including the SM and BSM contributions with the interference term. 
Here, we choose the strong $p_T(e)$ of 100 GeV to reject the major part of the background
coming from the $Z$-resonance. To capture various detector effects, we use the ILC detector 
card which is available in the Delphes package. Finally, we isolate the signal from the left-over SM background 
by applying the following stronger $p_T$ cuts on the outgoing electrons
\begin{align}
p_T(e_1),p_T(e_2) > 200~\textrm{GeV}
\end{align}
Note that, in this analysis, we have not really optimized the above $p_T$ cut to obtain maximum
sensitivity. But we found that this cut is good enough to obtain good significance and therefore we
keep the same $p_T$ value for all $M_{Z'}$ points. The definition of significance we use is given by 
$\sg=\mc{N}_S/\sqrt{\mc{N}_B}$, where $\mc{N}_S$ and $\mc{N}_B$ are the number of signal and background 
events respectively estimated for a particular luminosity.

In figure~\ref{fig:significance}, we show $2\sg$, $3\sg$ and $5\sg$ confidence level contours in the $M_{Z'}-\kp$ plane at the 0.5 TeV ILC with 100 fb$^{-1}$ integrated luminosity. One can see that a $Z'$ with mass around 3 TeV with
$\kp\sim 0.2$ can be discovered (with $5\sg$ CL) at the ILC. For the same coupling, ILC can rule out (with $2\sg$ CL)
$M_{Z'}$ up to around 7 TeV.

\section{Summary and discussion}\label{sec:discussion}
In this paper we have considered the collider phenomenology of a minimal $ \mathrm{U}(1) $ extension of the SM where the anomaly cancellation is not explicit, but through the GS mechanism. However, such a mechanism necessarily invites terms in the Lagrangian which are unsuppressed by the high scale which the physics resides at, and this physics can, at least in principle, be probed at lower energies. Some interesting examples of such models include gauged baryon number $ B $, gauged lepton number $ L $, and gauged $ B+L $. In equation\room\eqref{eq:charge}, we have derived the linear combinations of $ B,\room{}L $ and $ Y $ which can be gauged in the minimal scenario. We calculate the branching ratios of the different decay modes of $ Z' $, including the $ Z'\rightarrow ZZ $ and $ Z'\rightarrow Z\gamma $ modes, which are are loop suppressed. The branching ratios to these signatures are in general quite small but there still exists parameter space where these processes could be observed. 

For some benchmark models we put exclusions on the parameter space $ \left(g_{z},M_{Z'}\right) $. From previous studies of anomaly free $ \mathrm{U}(1) $ theories it is expected that the parameter space is heavily constrained for low $Z'$ masses. For GS $ \mathrm{U}(1) $ extensions it is possible to realize a wider class of models than in the strict anomaly free setting. For example, as can be seen in figure\room\ref{sfig:B}, the gauged $ B $ model does not receive any constraints from dilepton bounds and is harder to rule out at the LHC. The gauged $ L $ model is even more free in this regard because there are no couplings to quarks, so the $ Z' $ cannot be produced through quark fusion, and thus the bounds from LHC are very weak. Such a $ Z' $ can in principle be discovered at an $ e^{+} e^{-} $ collider such as the ILC.

The $Z' \rightarrow Z \gamma$ and $Z'\rightarrow Z Z$ processes are interesting since they receive contributions from the GS terms. These contributions are unfortunately loop suppressed and quite elusive at the LHC. By performing a random scan of the parameter space (shown in figure\room\ref{fig:scanHeatZZ}) we find that it is possible to have relatively large cross sections for the $ZZ$ and $Z\gamma$ channels, of order $1$~fb. The cross sections are too small to be detected by the current data and the possible detection of these processes is necessarily postponed to the HL-LHC. However, it can be expected that if a $Z'$ could be detected at the LHC it would most likely first be seen in the dijet or lepton channels after which the exact nature of the $Z'$ could be determined. 

Alternatively, lepton colliders provide stronger bounds than LHC for $Z'$ bosons that couple weakly to quarks. Naively it can be expected that the energy of lepton colliders is not large enough to resonantly produce $Z'$ bosons. Even though the chances for direct detection of $Z'$ bosons is slim at lepton colliders, the interference between $Z'$ bosons and SM processes makes it possible to indirectly probe the $Z'$ bosons, even for relatively large $Z'$ masses. Future lepton colliders are thus a good choice for indirectly studying quarkphobic $Z'$ models since they would be hard to detect at the LHC.

In conclusion, it is possible to relax some collider constraints on $Z'$ bosons in $ \mathrm{U}(1) $ extensions if the theories are extended with a GS mechanism. Furthermore, if a $ Z' $ with anomalous couplings, such as couplings proportional to $ B $, is discovered, then it is plausible to probe its GS nature at higher luminosity colliders via the $Z' \rightarrow Z \gamma$ and $Z'\rightarrow Z Z$ processes.
\section*{Acknowledgments}
This work was supported by the Swedish Research Council (contract 621-2011-5107) and the Carl Trygger Foundation (contract CTS-14:206).

\appendix
\section{Conventions}\label{app:ssb}
%\tcr{Formulea are different for the case ($M_z,M_Y\neq 0$ and $z_H=0$ or $z_H\neq 0$ hep-ph/0503208). 
%Maybe we can show them here.}
We consider the spontaneous symmetry breaking of $\mathrm{U}(1)_z$ through a St\"{u}ckelberg mechanism, as described in subsection\room\ref{ssec:GS}. The breaking will be analogous to a complex singlet acquiring a VEV and we hence skip the details (see\room\cite{Ekstedt:2016wyi}). After symmetry breaking, the photon field $A^{\mu}$ remains massless, while the other two physical fields $Z$ and $Z'$ acquire masses which are given by

\begin{equation}\label{eq:ZZ'mass}
M_{Z,Z'}=\frac{g v_H}{2 c_w}\left[\frac{1}{2}\lt\{(r+z_H^2)t_z^2c_w^2+1\rt\} \mp \frac{z_H t_zc_w}{\sin2\theta'}\right]^\frac{1}{2} ,
\end{equation}
where $t_z \equiv g_z/g$; $\tan\theta_w \equiv g'/g$ defines the Weinberg angle, and the parameter $r\equiv (2 M)^2/v_H^2$ is given in terms of the St\"{u}ckelberg  scale, M, and the Higgs doublet vev, $v_H$. The mixing angle $ \theta' $ satisfies
\begin{equation}\label{eq:mixang}
\theta'=\frac{1}{2}\arcsin\left(\frac{2 z_H t_z c_w}{\sqrt{\lt[2 z_H t_z c_w\rt]^2+\lt[(r+z_H^2)t_z^2c_w^2-1\rt]^2}}\right) .
\end{equation}
It is important to note that there can only be $Z\leftrightarrow Z'$ mixing at tree level if $ z_{H}\neq 0 $. The scale $ M $ can be written in terms of the other parameters as
\begin{equation}\label{eq:scale}
M^2=\frac{1}{4}v_H^2 {A}(M_{Z'}) \frac{ \left\{{A}(M_{Z'})-2-2 c_w^2 t_z^2 z_H^2\right\}}{2c_w^2 t_z^2\left\{{A}(M_{Z'})-2\right\}},
\end{equation}
where $ A(M_{Z'})\equiv 8 c_w^2 M_{Z'}^2/(g^2 v_H^2) $. Note that in the limit $ M_{Z'}\rightarrow\infty $, $ M\rightarrow M_{Z'}/g_{z} $.

If $\theta'\neq0$ the St\"{u}ckelberg axion will mix with the Goldstone bosons coming from the Higgs doublet. The mixing angle, $\theta_G$, is given by
\begin{align*}
\tan \theta_G=\tan \theta' ~\frac{M_{Z'}}{M_Z}.
\end{align*}

For the covariant derivative and gauge charges we use the convention
\begin{align}
\mathcal{D}^\mu=\left(\partial^\mu -i g W^\mu ~T^3_i -i Y_i~\frac{g'}{2}B_Y^\mu -i z_i ~\frac{g_z}{2}B_z^\mu\right).
\end{align}

\section{Loop amplitudes}\label{app:Loops}

\subsection{Rosenberg parametrization}\label{app:Schouten}

As an explicit example of how to rewrite triple-vector boson amplitudes in the Rosenberg parametrization, consider the $Z' Z Z $ amplitude
\begin{align*}
\Gamma^{Z' Z Z}_{\rho \mu \nu}(r,p,q) &= A \left( \epsilon[\mu ,\nu,p,q]q^\rho+\epsilon[\mu ,\nu,p,q]p^\rho\right)
\\&+ B\left(\epsilon[\mu,\nu, \rho,q] - \epsilon[\mu,\nu, \rho,p]\right)
\\& + C\left( \epsilon[\nu,\rho,p,q]q^\mu-\epsilon[\nu,\rho,p,q]p^\mu+\left(\mu\leftrightarrow\nu\right)\right)
%\\& +C \left(\epsilon[\mu,\rho,p,q]q^\nu- \epsilon[\mu,\rho,p,q]p^\nu\right).
\end{align*}
The Schouten-identity for a general 4-momentum $ P $ reads
\begin{align}\label{eq:schouten}
\epsilon_{\mu \nu \tau \sigma}P_\rho+\left(\text{cyclic permutations}\right)=0.
\end{align}
Taking $P=p$ and contracting with $p^\tau,q^\sigma$ the Schouten identity takes the form
\begin{align}
\epsilon[\mu,\nu,p,q]p_\rho=\epsilon[\nu,\rho,p,q]p_\mu+\epsilon[\mu,\rho,p,q]p_\nu+\epsilon[\mu,\nu,\rho,q]p^2-\epsilon[\mu,\nu,\tau,p]p \cdot q,
\end{align}
and similarly for $P=q$. The relation above enables us to remove the A-type terms in $\Gamma^{Z' V_1 V_2}_{\rho \mu \nu}(r,p,q)$ amplitudes, and redistribute them over the remaining Lorentz structures, which simply gives the familiar Rosenberg parametrization of the amplitude.

\subsection{General loop amplitude}

The generic amplitude from a fermion shown in fig.\room\ref{fig:triangle} is given by
\begin{align*}
A^{\rho \mu \nu}(r,p,q)=\int_{l} &\frac{\room\mathrm{Tr}\left[\left(\slashed{l}-\slashed{p}+m\right)\gamma^{\mu} G_1 \left(\slashed{l}+m\right)\gamma^{\nu} G_2 \left(\slashed{l}+\slashed{q}+m\right)\gamma^{\rho} G_3) \right]}{((l-p)^2-m^2)(l^2-m^2)((l+q)^2-m^2)}
\\&+\left(\mu\leftrightarrow \nu, p\leftrightarrow q\right),
\end{align*}
where $ r=p+q $ and the couplings $G_i$ are given in terms of the left-right projectors and couplings as 
\begin{align}
&G_1=\left(g_L^{1} P_L+g_R^{1} P_R\right),
\\&G_2=\left(g_L^{2} P_L+g_R^{2} P_R\right),
\\&G_3=\left(g_L^{3} P_L+g_R^{3} P_R\right).
\end{align}
In this paper we are only considering $Z'$ decays and subsequently we will always take $G_3=G_{Z'}$. Note that the Bose symmetry between particles 1 and 2 is ensured if $G_1=G_2$, which is the case for the $Z' \rightarrow Z Z$ process. The above amplitude is often equivalently parametrized in terms of axial/vector-couplings in the literature -- we here opt for the left-right parametrization out of convenience.

Suppressing the Lorentz indices, the amplitude $A$ can be decomposed as
\begin{align}
A(r,p,q)&=g_L^{1} g_L^{2}g_L^{3}A_{LLL}+g_L^{1} g_L^{2}g_R^{3}A_{LLR}+g_L^{1} g_R^{2}g_L^{3}A_{LRL}+\ldots+g_R^{1} g_R^{2}g_R^{3}A_{RRR},
\end{align}
where $A_{IJK}$ denotes $A(r,p,q)$ with $G_1,~ G_2,~G_3$ replaced with $I,~J,~K$. The anomalous terms reside in $g_L^{1} g_L^{2}g_L^{3} A_{LLL}+g_R^{1} g_R^{2}g_R^{3} A_{RRR}$, and it turns out\footnote{This easiest to see by expanding the amplitudes in vector and axial amplitudes.} that $A_{RRR}=-A_{LLL}$, such that the anomalous terms factorize as $\left(g_L^{1} g_L^{2}g_L^{3}-g_R^{1} g_R^{2}g_R^{3}\right)A_{LLL}$. All remaining terms vanish in the limit $m\rightarrow 0$.

While the above amplitude is finite, there are divergences that need to be regularized before they cancel. A popular method of regularizing triangle diagrams is to use a UV-cutoff, as is for example done in \cite{Anastasopoulos:2008jt}. We find it more practical to use dimensional regularization (DR). However, as is widely known, naive DR with an anti-commuting $\gamma_5$ is inconsistent and hence we use the consistent BMHV-scheme (Breitenlohner-Maison-'t~Hooft-Veltman)\room\cite{tHooft:1972tcz,Breitenlohner:1977hr}. This scheme has the property that the $A_{LLL},A_{RRR}$ anomalies are automatically distributed symmetrically over all vector bosons, \textit{i.e}., the symmetric anomaly scheme is automatically built in. A word of caution: the BMHV scheme, while being consistent, is notorious for breaking BRST-invariance. This necessitates the introduction of gauge-variant counter-terms -- fortunately this is straightforward for the triple gauge boson processes of interest in this paper. For a more detailed account of the subtleties of the BMHV scheme we refer the interested reader to\room\cite{Tsai:2010aq,SanchezRuiz:2002xc,Martin:1999cc} and references therein.

\subsection{$A_{IJK}$ amplitudes}\label{app:Genamplitudes}

We present all sub-amplitudes as functions of $p^2,~q^2,\room r^2$ and the fermion mass $ m $ and we assume that all particles are outgoing:~$r+p+q=0$. The amplitudes are given in terms of the triangle Passarino-Veltman scalar function\room\cite{Passarino:1978jh,tHooft:1978jhc}. For convenience we define the functions
\begin{align}
L_i &\equiv \sqrt{p_i^2(p_i^2-4m^2)}\log\left[\frac{2m^2-p_i^2+\sqrt{p_i^2(p_i^2-4m^2)}}{2m^2}\right],\\
\Delta &\equiv p^4+q^4+r^4-2(r^2q^2-2p^2 q^2-2p^2 r^2),
\end{align}
and denote the triangle Passarino-Veltman function $C_0$ as $C\equiv C_0(p^2,q^2,r^2,m^2,m^2,m^2)$, with $i=p,q,r$. All the amplitudes are accompanied by an overall factor $\frac{m^ 2}{\Delta 4 \pi^2 }$ which is left implicit. The coefficient of the various Lorentz structures of the amplitudes can then be brought to the form 
\begin{itemize}
\item $A_{RLL}$:
\begin{itemize}
 \item $\epsilon[\mu,\nu,\rho,q]$: 
	\begin{align*}
	C\left\lbrace p^2\left(q^2+r^2\right)-\left(q^2-r^2\right)^2\right\rbrace
	 +2 L_p-L_q-L_r+L_q \frac{r^2-p^2}{q^2}+L_r \frac{q^2-p^2}{r^2}
	\end{align*}

\item $\epsilon[\mu,\nu,\rho,p]:$ 
	\begin{align*}
	C\left\lbrace q^2 \left(r^2+p^2\right)-q^4\right\rbrace
	+L_p+L_r-2 L_q+L_p\frac{q^2-r^2}{p^2}+L_r\frac{q^2-p^2}{r^2}.
	\end{align*}
\end{itemize}

\item $A_{LRL}$:
\begin{itemize}
 \item $\epsilon[\mu,\nu,\rho,q]:$ 
	\begin{align*}
	C\left\lbrace p^4-p^2\left(q^2+r^2\right)\right\rbrace
	 +2 L_p-L_q-L_r+L_q \frac{r^2-p^2}{q^2}+L_r \frac{q^2-p^2}{r^2}
	\end{align*}

\item $\epsilon[\mu,\nu,\rho,p]:$ 
	\begin{align*}
	C\left\lbrace \left(p^2-r^2\right)^2-q^2\left(p^2+r^2\right)\right\rbrace
	+L_p+L_r-2 L_q+L_p\frac{q^2-r^2}{p^2}+L_r\frac{q^2-p^2}{r^2}.
	\end{align*}
\end{itemize}

\item $A_{LLR}$:
\begin{itemize}
 \item $\epsilon[\mu,\nu,\rho,q]:$ 
	\begin{align*}
	C\left\lbrace p^2\left(p^2-q^2-r^2\right)\right\rbrace
	 +2 L_p-L_q-L_r+L_q\frac{r^2-p^2}{q^2}+L_r\frac{q^2-p^2}{r^2}.
	\end{align*}

\item $\epsilon[\mu,\nu,\rho,p]=-\epsilon[\mu,\nu,\rho,q]|_{p\leftrightarrow q}$
\end{itemize}

\end{itemize}

The amplitude $A_{RRR}=-A_{LLL}$ is considerably messier and it contain terms that are finite when $m^2\rightarrow 0$. Hence we only factor our an overall factor $\frac{1}{\Delta 4 \pi^2 }$ for this amplitude.
\begin{itemize}
	\item $A_{RRR}$:
	\begin{itemize}
		\item $\epsilon[\mu,\nu,p,q]q^\rho:$ 
		\begin{align*}
		&C\left(m^2(r^2-3p^2-q^2)+\frac{p^2(q^2(-3r^2p^2-2p^4+3r^4)+4p^2q^4-r^2(r^2-p^2)^2-2q^6}{\Delta(r,p,q)}\right)
		\\&	-\frac{3}{2\Delta(r,p,q)}\left(p^2-q^2+r^2)(p^2+3q^2-r^2)\right)L_p
		\\&+\frac{1}{2q^2\Delta(r,p,q)}\left((-p^2(7q^4+r^4)-p^4(r^2-8q^2)+p^6+(r^2-2q^2)(r^2-q^2)^2\right) L_q
		\\&+\frac{1}{2r^2 \Delta(r,p,q)}\left(-2p^2(-6r^2q^2+q^4+r^2)+p^4(r^2-2q^2)+2p^6-(r^2-2q^2)(r^2-q^2)^2\right)L_r,
		\end{align*}%
		\item $\epsilon[\mu,\nu,\rho,q]:$ 
		\begin{align*}
		&\frac{1}{2 }\left(p^2(m^2-r^2)-m^2(r^2-q^2)^2+r^2p^2\right) C
		\\& +\frac{1}{4}\left(-p^2+q^2+8m^2-3r^2\right) L_p
		\\& +\frac{1}{4 q^2}\left(-p^2(q^2+4m^2)+(4m^2-r^2)(r^2-q^2)+p^4\right) L_q
		\\& +\frac{1}{ r^2}\left(p^2(3r^2-4m^2)-(4m^2-r^2)(r^2-q^2)\right) L_r
		\\&-\frac{(p^4+(q^2-r^2)^2-2p^2(q^2+r^2))}{12}
		\end{align*}%
		\item $\epsilon[\nu,\rho,p,q]q^\mu:$
		\begin{align*}
		&-C\left(m^2 (p^2+r^2-q^2)+\frac{r^2p^2(q^2(p^2+r^2)+(r^2-p^2)^2-2q^4)}{\Delta(r,p,q)}\right)
		%\\&-\frac{1}{2\Delta(r,p,q)}\left(p^2(4r^2-2q^2)+p^4+4r^2q^2+q^4-5r^4\right)L_q
		\\&-\frac{1}{2\Delta(r,p,q)}\left(4r^{2}(p^{2}+q^{2}) +(p^{2}-q^{2})^{2}-5 r^{4}\right)L_p
		\\&+\frac{1}{2 q^2\Delta(r,p,q)}\left(p^2(8r^2 q^2+q^4-r^4)-p^4(2q^2+r^2)+p^6+r^2(r^2-q^2	)^2\right)L_q
		\\&-\frac{1}{2 \Delta(r,p,q)}\left(4p^2(q^2+r^2)-5p^4+(r^2-q^2)^2\right)L_r
		\\&-\frac{(p^2-q^2+r^2)(p^4+(q^2-r^2)^2-2p^2(q^2+r^2))}{2 \Delta(r,p,q)}.
		\end{align*}		
	\end{itemize}
\end{itemize}
The remaining Lorentz structures can be obtained from the symmetry relations
\begin{align*}
	\epsilon[\mu,\nu,p,q]q^\rho&=\epsilon[\mu,\nu,p,q]p^\rho|_{p\leftrightarrow q},\\
	\epsilon[\mu,\nu,\rho,p]&=-\epsilon[\mu,\nu,\rho,q]^\rho|_{p\leftrightarrow q},\\
	\epsilon[\nu,\rho,p,q]p^\mu&=-\epsilon[\nu,\rho,p,q]q^\mu|_{p\leftrightarrow q},\\
	\epsilon[\mu,\rho,p,q]p^\nu&=\epsilon[\nu,\rho,p,q]q^\mu|_{p\leftrightarrow q},\\
	\epsilon[\mu,\rho,p,q]q^\nu&=\epsilon[\nu,\rho,p,q]q^\mu|_{p\leftrightarrow q}.
\end{align*}
Note that terms of the form $-\frac{p^2-q^2+r^2}{8\pi^2\Delta(r,p,q)}$ combine when contracted with external momenta to give the right anomaly terms.

The remaining amplitudes can be obtained from the relations\footnote{These relations follows directly from that the $A_{VVV}$ and $A_{VAA}$ amplitudes vanish.}
\begin{align*}
A_{RRL}=&-A_{LLR},
\\A_{RLR}=&-A_{LRL},
\\A_{LRR}=&-A_{RLL},
\\A_{LLL}=&-A_{RRR}.
\end{align*}
It should be noted that the sub-amplitudes above are very general and greatly simplify when studying specific amplitudes. The amplitudes given above are valid for all fermion masses $ m $, including $m^2=0$, as long as all external legs have time-like momenta. If $ m^{2}=0 $ and one of the external legs is light-like, this case has to be treated with care and we refer the interested reader to the discussion in section \ref{app:massless}.

\subsection{Massless limit}\label{app:massless}

The $A_{IJK}$ amplitudes given in section \ref{app:Genamplitudes} seem to diverge for massless fermions when one of the external legs is light-like. However, these divergences will always drop out in the end, if care is taken when performing the massless limit. The way this works can be a bit subtle and this section will be dedicated to this issue.

In the massless fermion limit only the $\Gamma_{LLL}=-\Gamma_{RRR}$ sub-amplitudes are non-vanishing, but a direct application of the formulas in section \ref{app:Genamplitudes} will result in a divergent and ill-defined result. The origin of this divergence is that the limit $m\rightarrow 0$ invalidates the scalar integral decomposition if $p_\gamma^2=q^2=0$ (This is not a problem for the $Z'\rightarrow Z Z$ amplitude since the Z boson is massive). The proper way to treat the massless limit is to consider $q^2\neq 0$ and take the limit $\frac{m^2}{q^2}\rightarrow 0$,\footnote{We refer the interested reader to \cite{Carlitz:1988ab} for an in-depth discussion} or in practice take $q^2 \neq 0$ and set $m=0$ at the beginning of the calculation. Taking the limit $q^2\rightarrow 0$, the $\Gamma_{RRR}$ (and equivalently $\Gamma_{LLL}$) amplitude will contain potentially divergent terms of the type $\frac{1}{\epsilon_{IR}}$ and $\log\left(\frac{\mu^2}{q^2}\right)$, which naively seem to be ill-defined in the limit where the photon goes on-shell. The amplitude is however rendered finite, since all divergent terms are proportional to the tensor structure
\begin{align}
\left(-\epsilon[\mu,\nu,\rho,q] (p\cdot q)+\epsilon[\nu,\rho,p,q]q^\mu+\epsilon[\mu,\rho,p,q]q^\nu+\epsilon[\mu,\nu,p,q]q^\rho\right),
\end{align} 
which after use of the Schouten identity\room\ref{eq:schouten} reduces to
\begin{align}
 2\epsilon[\mu,\rho,p,q]q^\nu.
\end{align}
The term $\epsilon[\mu,\rho,p,q]q^\nu$ will drop out of any calculation in the limit $q^2 \rightarrow 0$ and any remaining terms are finite. 
After the proper limits have been taken the surviving terms in the $\Gamma_{RRR}$ amplitude are

\begin{align*}
\Gamma_{RRR} =&- \epsilon[\mu,\nu,\rho,q]\frac{\left(p^2-r^2\right)^2-3 p^2 \log \left(\frac{p^2}{r^2}\right) \left(-r^2 \log\left(\frac{p^2}{r^2}\right)+p^2+3 r^2\right)}{48 \pi ^2 \left(p^2-r^2\right)^2}
\\&+\epsilon[\mu,\nu,\rho,p]	\frac{3 r^2 \log \left(\frac{p^2}{r^2}\right)+p^2-r^2}{48 \pi ^2 \left(p^2-r^2\right)}
\\&- \epsilon[\mu,\rho,p,q]p^\nu \frac{p^2-r^2 \left(\log \left(\frac{p^2}{r^2}\right)+1\right)}{8 \pi ^2
   \left(p^2-r^2\right)^2}
   \\&+
   \epsilon[\mu,\nu,p,q]q^\rho\frac{-6 p^2 r^2 \left(\log \left(\frac{p^2}{r^2}\right)-4\right) \left(\log
   \left(\frac{p^2}{r^2}\right)+1\right)+18 p^4 \left(\log
   \left(\frac{p^2}{r^2}\right)-1\right)-6 r^4}{48 \pi ^2 \left(p^2-r^2\right)^3}
   \\&+\epsilon[\mu,\nu,p,q]q^\rho\frac{\left(2 p^2-r^2\right) \log \left(\frac{p^2}{r^2}\right)-p^2+r^2}{8 \pi ^2
   \left(p^2-r^2\right)^2}
   \\&-\epsilon[\nu,\rho,p,q]p^\mu	\frac{p^2-r^2 \left(\log \left(\frac{p^2}{r^2}\right)+1\right)}{8 \pi ^2
   \left(p^2-r^2\right)^2}
   \\&+\epsilon[\nu,\rho,p,q]q^\mu\frac{-p^4+p^2 \log \left(\frac{p^2}{r^2}\right) \left(-r^2 \log
   \left(\frac{p^2}{r^2}\right)+p^2+5 r^2\right)+r^4}{8 \pi ^2 \left(p^2-r^2\right)^3}.
\end{align*}

If in analogy with the $Z'\rightarrow Z \gamma$ process we decompose $\Gamma_{RRR}$ as
\begin{align*}
\Gamma_{RRR} =& A_1~ \epsilon[\mu ,\nu,p,q]q^\rho+ A_2~  \epsilon[\mu ,\nu,p,q]p^\rho\nonumber
\\&+ A_3 ~\epsilon[\mu,\nu, \rho,q]+ A_4 ~\epsilon[\mu,\nu, \rho,p]\nonumber
\\& + A_5 ~\epsilon[\nu,\rho,p,q]q^\mu+ A_6 ~\epsilon[\nu,\rho,p,q]p^\mu\nonumber
\\& + A_7 ~\epsilon[\mu,\rho,p,q]q^\nu+ A_8 ~\epsilon[\mu,\rho,p,q]p^\nu.
\end{align*}
Using the Schouten identity and the transversality of the external polarization tensors it is straightforward to rewrite the amplitude as

\begin{align*}
\Gamma_{RRR} =& B_1 ~\epsilon[\mu,\nu, \rho,q]+ B_2 ~\epsilon[\mu,\nu, \rho,p]\nonumber
\\& + B_3 ~\epsilon[\nu,\rho,p,q]q^\mu+ B_4 ~\epsilon[\mu,\rho,p,q]p^\nu,
\end{align*}
where
\begin{align*}
&B_1=\frac{5 p^4-p^2 r^2 \left(9 \log \left(\frac{p^2}{r^2}\right)+4\right)+r^4 \left(3 \log
   \left(\frac{p^2}{r^2}\right)-1\right)}{48 \pi ^2 \left(p^2-r^2\right)^2},
\\&B_2=\frac{r^2 \left(3 \log \left(\frac{p^2}{r^2}\right)-1\right)+p^2}{48 \pi ^2
   \left(p^2-r^2\right)},
   \\& B_3=\frac{p^2-r^2 \left(\log \left(\frac{p^2}{r^2}\right)+1\right)}{8 \pi ^2
   \left(p^2-r^2\right)^2},
   \\&B_4=\frac{r^2 \left(\log \left(\frac{p^2}{r^2}\right)+1\right)-p^2}{8 \pi ^2
   \left(p^2-r^2\right)^2}.
\end{align*}

For the $Z' \rightarrow Z\gamma$ process the net effect of the above considerations is that the form factor $F_1$ take the simple form
\begin{align}
F_1=\iu \frac{(g^Z_R g^{Z'}_R-g^Z_L g^{Z'}_L)Q \left[M_Z^2-M_{Z'}^2+M_{Z'}^2\log\left(\frac{M_{Z'}^2}{M_{Z}^2}\right)\right]}{8\pi^2\left(M_{Z}^2-M_{Z'}^2\right)^2}.
\end{align}

\bibliographystyle{myBibStyle} 
\bibliography{Bibliography_zpGS}

\begin{thebibliography}{99}
\providecommand{\url}[1]{\texttt{#1}}
\providecommand{\urlprefix}{URL }
\providecommand{\eprint}[2][]{\url{#2}}

\bibitem{Appelquist:2002mw}
T.~Appelquist, B.~A. Dobrescu, and A.~R. Hopper.
\newblock \emph{Nonexotic neutral gauge bosons}.
\newblock \emph{Phys. Rev.}, D68:035012, 2003.
\newblock \eprint{arXiv:hep-ph/0212073}.

\bibitem{Ekstedt:2016wyi}
A.~Ekstedt, R.~Enberg, G.~Ingelman, J.~Löfgren, and T.~Mandal.
\newblock \emph{{Constraining minimal anomaly free $\mathrm{U}(1)$ extensions
  of the Standard Model}}.
\newblock \emph{JHEP}, 11:071, 2016.
\newblock \eprint{arXiv:1605.04855}.

\bibitem{Green:1984sg}
M.~B. Green and J.~H. Schwarz.
\newblock \emph{{Anomaly cancellation in supersymmetric $D=10$ gauge theory and
  superstring theory}}.
\newblock \emph{Phys. Lett.}, B149:117, 1984.

\bibitem{Antoniadis:2010zza}
I.~Antoniadis, A.~Boyarsky, and O.~Ruchayskiy.
\newblock \emph{{Anomaly driven signatures of extra U(1)'s}}.
\newblock \emph{AIP Conf. Proc.}, 1200:64, 2010.

\bibitem{Anastasopoulos:2006cz}
P.~Anastasopoulos, M.~Bianchi, E.~Dudas, and E.~Kiritsis.
\newblock \emph{{Anomalies, anomalous U(1)'s and generalized Chern-Simons
  terms}}.
\newblock \emph{JHEP}, 11:057, 2006.
\newblock \eprint{arXiv:0605225}.

\bibitem{Anastasopoulos:2008jt}
P.~Anastasopoulos, F.~Fucito, A.~Lionetto, G.~Pradisi, A.~Racioppi, and Y.~S.
  Stanev.
\newblock \emph{{Minimal anomalous $\mathrm{U}(1)'$ extension of the MSSM}}.
\newblock \emph{Phys. Rev.}, D78:085014, 2008.
\newblock \eprint{arXiv:0804.1156}.

\bibitem{Ismail:2017ulg}
A.~Ismail, A.~Katz, and D.~Racco.
\newblock \emph{{On dark matter interactions with the Standard Model through an
  anomalous $Z'$}}.
\newblock 2017.
\newblock \eprint{arXiv:1707.00709}.

\bibitem{Dror:2017ehi}
J.~A. Dror, R.~Lasenby, and M.~Pospelov.
\newblock \emph{{New constraints on light vectors coupled to anomalous
  currents}}.
\newblock 2017.
\newblock \eprint{arXiv:1705.06726}.

\bibitem{Dror:2017nsg}
J.~A. Dror, R.~Lasenby, and M.~Pospelov.
\newblock \emph{{Dark forces coupled to non-conserved currents}}.
\newblock \emph{Phys. Rev.}, D96:075036, 2017.
\newblock \eprint{arXiv:1707.01503}.

\bibitem{Ismail:2017fgq}
A.~Ismail and A.~Katz.
\newblock \emph{{Anomalous $Z'$ and diboson resonances at the LHC}}.
\newblock 2017.
\newblock \eprint{arXiv:1712.01840}.

\bibitem{WeinbergII}
S.~Weinberg.
\newblock \emph{The quantum theory of fields: Vol. 2, Modern applications}.
\newblock Cambridge Univ. Press, Cambridge, 1996.

\bibitem{Adler:1969er}
S.~L. Adler and W.~A. Bardeen.
\newblock \emph{{Absence of higher order corrections in the anomalous axial
  vector divergence equation}}.
\newblock \emph{Phys. Rev.}, 182:1517, 1969.

\bibitem{Adler:1969gk}
S.~L. Adler.
\newblock \emph{{Axial vector vertex in spinor electrodynamics}}.
\newblock \emph{Phys. Rev.}, 177:2426, 1969.

\bibitem{Fujikawa:1979ay}
K.~Fujikawa.
\newblock \emph{{Path integral measure for gauge invariant fermion theories}}.
\newblock \emph{Phys. Rev. Lett.}, 42:1195, 1979.

\bibitem{Fujikawa:1980eg}
K.~Fujikawa.
\newblock \emph{{Path integral for gauge theories with fermions}}.
\newblock \emph{Phys. Rev.}, D21:2848, 1980.
\newblock [Erratum: Phys. Rev.D22,1499(1980)].

\bibitem{Bilal:2008qx}
A.~Bilal.
\newblock \emph{{Lectures on anomalies}}.
\newblock 2008.
\newblock \eprint{arXiv:0802.0634}.

\bibitem{Stueckelberg:1900zz}
E.~C.~G. Stueckelberg.
\newblock \emph{{Interaction energy in electrodynamics and in the field theory
  of nuclear forces}}.
\newblock \emph{Helv. Phys. Acta}, 11:225, 1938.

\bibitem{Peccei:1977hh}
R.~D. Peccei and H.~R. Quinn.
\newblock \emph{{CP conservation in the presence of instantons}}.
\newblock \emph{Phys. Rev. Lett.}, 38:1440, 1977.

\bibitem{Carone:1995pu}
C.~D. Carone and H.~Murayama.
\newblock \emph{{Realistic models with a light $\mathrm{U}(1)$ gauge boson coupled to
  baryon number}}.
\newblock \emph{Phys. Rev.}, D52:484, 1995.
\newblock \eprint{arXiv:hep-ph/9501220}.

\bibitem{Aranda:2014zta}
A.~Aranda, E.~Jiménez, and C.~A. Vaquera-Araujo.
\newblock \emph{{Electroweak phase transition in a model with gauged lepton
  number}}.
\newblock \emph{JHEP}, 01:070, 2015.
\newblock \eprint{arXiv:1410.7508}.

\bibitem{Mohapatra:1982xz}
R.~N. Mohapatra and G.~Senjanovic.
\newblock \emph{{Spontaneous breaking of global $B-L$ symmetry and 
     matter-antimatter oscillations in grand unified theories}}.
\newblock \emph{Phys. Rev.}, D27:254, 1983.

\bibitem{Erler:1999ub}
J.~Erler and P.~Langacker.
\newblock \emph{{Constraints on extended neutral gauge structures}}.
\newblock \emph{Phys. Lett.}, B456:68, 1999.
\newblock \eprint{arXiv:hep-ph/9903476}.

\bibitem{DeRujula:1980qc}
A.~De~Rujula, H.~Georgi, and S.~L. Glashow.
\newblock \emph{{Flavor goniometry by proton decay}}.
\newblock \emph{Phys. Rev. Lett.}, 45:413, 1980.

\bibitem{Alwall:2014hca}
J.~Alwall, R.~Frederix, S.~Frixione, V.~Hirschi, F.~Maltoni, O.~Mattelaer,
  H.~S. Shao, T.~Stelzer, P.~Torrielli, and M.~Zaro.
\newblock \emph{{The automated computation of tree-level and next-to-leading
  order differential cross sections, and their matching to parton shower
  simulations}}.
\newblock \emph{JHEP}, 07:079, 2014.
\newblock \eprint{arXiv:1405.0301}.

\bibitem{Shtabovenko:2016sxi}
V.~Shtabovenko, R.~Mertig, and F.~Orellana.
\newblock \emph{{New Developments in FeynCalc 9.0}}.
\newblock \emph{Comput. Phys. Commun.}, 207:432, 2016.
\newblock \eprint{arXiv:1601.01167}.

\bibitem{Mertig:1990an}
R.~Mertig, M.~Bohm, and A.~Denner.
\newblock \emph{{FeynCalc: Computer algebraic calculation of Feynman
  amplitudes}}.
\newblock \emph{Comput. Phys. Commun.}, 64:345, 1991.

\bibitem{Alloul:2013bka}
A.~Alloul, N.~D. Christensen, C.~Degrande, C.~Duhr, and B.~Fuks.
\newblock \emph{{FeynRules 2.0 - A complete toolbox for tree-level
  phenomenology}}.
\newblock \emph{Comput. Phys. Commun.}, 185:2250, 2014.
\newblock \eprint{arXiv:1310.1921}.

\bibitem{Kublbeck:1990xc}
J.~Kublbeck, M.~Bohm, and A.~Denner.
\newblock \emph{{FeynArts: Computer algebraic generation of Feynman graphs and
  amplitudes}}.
\newblock \emph{Comput. Phys. Commun.}, 60:165, 1990.

\bibitem{Hahn:2000kx}
T.~Hahn.
\newblock \emph{{Generating Feynman diagrams and amplitudes with FeynArts 3}}.
\newblock \emph{Comput. Phys. Commun.}, 140:418, 2001.
\newblock \eprint{arXiv:hep-ph/0012260}.

\bibitem{Patel:2015tea}
H.~H. Patel.
\newblock \emph{{Package-X: A Mathematica package for the analytic calculation
  of one-loop integrals}}.
\newblock \emph{Comput. Phys. Commun.}, 197:276, 2015.
\newblock \eprint{arXiv:1503.01469}.

\bibitem{Shtabovenko:2016whf}
V.~Shtabovenko.
\newblock \emph{{FeynHelpers: Connecting FeynCalc to FIRE and Package-X}}.
\newblock \emph{Comput. Phys. Commun.}, 218:48, 2017.
\newblock \eprint{arXiv:1611.06793}.

\bibitem{Rosenberg:1962pp}
L.~Rosenberg.
\newblock \emph{{Electromagnetic interactions of neutrinos}}.
\newblock \emph{Phys. Rev.}, 129:2786, 1963.

\bibitem{Ball:2012cx}
R.~D. Ball et~al.
\newblock \emph{{Parton distributions with LHC data}}.
\newblock \emph{Nucl. Phys.}, B867:244, 2013.
\newblock \eprint{arXiv:1207.1303}.

\bibitem{Gumus:2006mxa}
K.~Gumus, N.~Akchurin, S.~Esen, and R.~M. Harris.
\newblock \emph{{CMS Sensitivity to Dijet Resonances}}.
\newblock CMS-NOTE-2006-070.
\newblock 2006.

\bibitem{Aaboud:2017buh}
M.~Aaboud et~al. (ATLAS).
\newblock \emph{{Search for new high-mass phenomena in the dilepton final state
  using 36.1 fb$^{-1}$ of proton-proton collision data at $\sqrt{s}$ = 13 TeV
  with the ATLAS detector}}.
\newblock 2017.
\newblock \eprint{arXiv:1707.02424}.

\bibitem{Aaboud:2017yvp}
M.~Aaboud et~al. (ATLAS).
\newblock \emph{{Search for new phenomena in dijet events using 37 fb$^{-1}$ of
  $pp$ collision data collected at $\sqrt{s}=$13 TeV with the ATLAS detector}}.
\newblock 2017.
\newblock \eprint{arXiv:1703.09127}.

\bibitem{CMS:2017xrr}
C.~Collaboration (CMS).
\newblock \emph{{Searches for dijet resonances in pp collisions at
  $\sqrt{s}=13~\mathrm{TeV}$ using data collected in 2016.}}
\newblock CMS-PAS-EXO-16-056.
\newblock 2017.

\bibitem{Olive:2016xmw}
C.~Patrignani et~al. (Particle Data Group).
\newblock \emph{{Review of Particle Physics}}.
\newblock \emph{Chin. Phys.}, C40(10):100001, 2016.

\bibitem{DelAguila:1993rw}
F.~Del~Aguila and M.~Cvetic.
\newblock \emph{{Diagnostic power of future colliders for $Z'$ couplings to
  quarks and leptons: $e^+ e^{-}$ versus $pp$ colliders}}.
\newblock \emph{Phys. Rev.}, D50:3158, 1994.
\newblock \eprint{arXiv:hep-ph/9312329}.

\bibitem{Godfrey:1994qk}
S.~Godfrey.
\newblock \emph{{Comparison of discovery limits for extra Z bosons at future
  colliders}}.
\newblock \emph{Phys. Rev.}, D51:1402, 1995.
\newblock \eprint{arXiv:hep-ph/9411237}.

\bibitem{DelAguila:1995fa}
F.~Del~Aguila, M.~Cvetic, and P.~Langacker.
\newblock \emph{{Reconstruction of the extended gauge structure from $Z'$
  observables at future colliders}}.
\newblock \emph{Phys. Rev.}, D52:37, 1995.
\newblock \eprint{arXiv:hep-ph/9501390}.

\bibitem{Godfrey:2005pm}
S.~Godfrey, P.~Kalyniak, and A.~Tomkins.
\newblock \emph{{Distinguishing between models with extra gauge bosons at the
  ILC}}.
\newblock In \emph{{Proceedings, 2005 International Linear Collider Physics and
  Detector Workshop and 2nd ILC Accelerator Workshop (Snowmass 2005)}}. 2005.
\newblock \eprint{arXiv:hep-ph/0511335}.

\bibitem{Osland:2009dp}
P.~Osland, A.~A. Pankov, and A.~V. Tsytrinov.
\newblock \emph{{Identification of extra neutral gauge bosons at the
  International Linear Collider}}.
\newblock \emph{Eur. Phys. J.}, C67:191, 2010.
\newblock \eprint{arXiv:0912.2806}.

\bibitem{Battaglia:2012ez}
M.~Battaglia, F.~Coradeschi, S.~De~Curtis, and D.~Dominici.
\newblock \emph{{Indirect sensitivity to heavy $Z'$ bosons at a multi-TeV $e^+
  e^{-}$ collider}}.
\newblock In \emph{{International Workshop on Future Linear Colliders (LCWS11)
  Granada, Spain, September 26-30, 2011}}. 2012.
\newblock \eprint{arXiv:1203.0416}.

\bibitem{Han:2013mra}
T.~Han, P.~Langacker, Z.~Liu, and L.-T. Wang.
\newblock \emph{{Diagnosis of a new neutral gauge boson at the LHC and ILC for
  Snowmass 2013}}.
\newblock 2013.
\newblock \eprint{arXiv:1308.2738}.

\bibitem{tHooft:1972tcz}
G.~'t~Hooft and M.~J.~G. Veltman.
\newblock \emph{{Regularization and renormalization of gauge fields}}.
\newblock \emph{Nucl. Phys.}, B44:189, 1972.

\bibitem{Breitenlohner:1977hr}
P.~Breitenlohner and D.~Maison.
\newblock \emph{{Dimensional renormalization and the action principle}}.
\newblock \emph{Commun. Math. Phys.}, 52:11, 1977.

\bibitem{Tsai:2010aq}
E.-C. Tsai.
\newblock \emph{{Maintaining gauge symmetry in renormalizing chiral gauge
  theories}}.
\newblock \emph{Phys. Rev.}, D83:065011, 2011.
\newblock \eprint{arXiv:1012.3501}.

\bibitem{SanchezRuiz:2002xc}
D.~Sanchez-Ruiz.
\newblock \emph{{BRS symmetry restoration of chiral Abelian Higgs-Kibble theory
  in dimensional renormalization with a nonanticommuting $\gamma^5$}}.
\newblock \emph{Phys. Rev.}, D68:025009, 2003.
\newblock \eprint{arXiv:hep-th/0209023}.

\bibitem{Martin:1999cc}
C.~P. Martin and D.~Sanchez-Ruiz.
\newblock \emph{{Action principles, restoration of BRS symmetry and the
  renormalization group equation for chiral nonAbelian gauge theories in
  dimensional renormalization with a nonanticommuting $\gamma^5$}}.
\newblock \emph{Nucl. Phys.}, B572:387, 2000.
\newblock \eprint{arXiv:hep-th/9905076}.

\bibitem{Passarino:1978jh}
G.~Passarino and M.~J.~G. Veltman.
\newblock \emph{{One loop corrections for $e^+ e^{-}$ annihilation into $\mu^+
  \mu^{-}$ in the Weinberg Model}}.
\newblock \emph{Nucl. Phys.}, B160:151, 1979.

\bibitem{tHooft:1978jhc}
G.~'t~Hooft and M.~J.~G. Veltman.
\newblock \emph{{Scalar one loop integrals}}.
\newblock \emph{Nucl. Phys.}, B153:365, 1979.

\bibitem{Carlitz:1988ab}
R.~D. Carlitz, J.~C. Collins, and A.~H. Mueller.
\newblock \emph{{The role of the axial anomaly in measuring spin dependent
  parton distributions}}.
\newblock \emph{Phys. Lett.}, B214:229, 1988.

\end{thebibliography}
\end{document}